\pdfoutput=1
\documentclass{aa}  

\usepackage{graphicx}
\usepackage{txfonts}
\usepackage{natbib}
\usepackage[draft]{hyperref}
\usepackage{xcolor}

\begin{document}

   \title{A young SNR illuminating nearby Molecular Clouds with cosmic rays}

   \author{Y. Cui \inst{1}, G. P\"uhlhofer \inst{1}, A. Santangelo \inst{1}
          }

   \institute{$^1$Institut f\"ur Astronomie und Astrophysik Universit\"at T\"ubingen \\
              \email{yudong.cui@astro.uni-tuebingen.de}        
             }

   \date{submitted on }

 \abstract
{
The Supernova Remnant (SNR) HESS J1731-347 displays strong non-thermal TeV $\gamma$-ray and X-ray emission, thus the object is at present time accelerating particles to very high energies. A distinctive feature of this young SNR is the nearby ($\sim 30\,\rm{pc}$ in projection) extended source HESS J1729-345, which is currently unidentified but is in spatial projection coinciding with known molecular clouds (MC). We model the SNR evolution to explore if the TeV emission from HESS J1729-345 can be explained as emission from runaway hadronic cosmic rays (CRs) that are illuminating these MCs.
The observational data of HESS J1729-345 and HESS J1731-347 can be reproduced using core-collapse SN models for HESS J1731-347. Starting with different progenitor stars and their pre-supernova environment, we model potential SNR evolution histories along with the CR acceleration in the SNR and the diffusion of the CRs. A simplified 3-dimensional structure of the MCs is introduced based on $\rm{^{12}CO}$ data of that region, adopting a distance of $3.2\,\rm{kpc}$ to the source. A Monte Carlo-based diffusion model for the escaping CRs is developed to deal with the inhomogeneous environment. The fast SNR forward shock speed as implied from the X-ray data can easily be explained when employing scenarios with progenitor star masses between $20\,\rm{M_\odot}$ and $25\,\rm{M_\odot}$, where the SNR shock is still expanding inside the main sequence (MS)-bubble at present time. The \rm{TeV} spectrum of HESS J1729-345 is satisfactorily fitted by the emission from the highest-energy CRs that have escaped the SNR, using a standard galactic CR diffusion coefficient in the inter-clump medium. The \rm{TeV} image of HESS J1729-345 can be explained with a reasonable 3-dimensional structure of MCs. The TeV emission from the SNR itself is dominated by leptonic emission in this model. We also explore scenarios where the shock is starting to encounter the dense MS progenitor wind bubble shell. The escaping hadronic CR hypothesis for the $\gamma$-ray emission of HESS J1729-345 can still hold, but our model can also in this case not easily account for the TeV emission from HESS J1731-347 in a hadronic scenario.
}
   \keywords{Astroparticle physics - Cosmic rays  -  Gamma-rays: ISM -  ISM: supernova remnants -   ISM: individual objects: HESS J1731-347
               }

   \maketitle

\section{Introduction}
Supernova Remnants (SNRs) are considered prime candidates for the main sources of Galactic Cosmic Rays (CRs), given the sustained kinetic energy input of this source class into the Galaxy and the well-known Fermi mechanism to convert this energy into CR particle acceleration. Substantial theoretical improvement has been achieved to support the idea that SNR shocks can indeed accelerate particles up to ``knee'' energies, i.e. up to $\sim$$10^{15}\,\mathrm{eV}$, employing the concept of fast amplification of magnetic fields upstream of SNR shocks through non-resonant streaming instability \citep{Bell2004,Zirakashvili2008a,Zirakashvili2008b}. Direct observational evidence for particle acceleration up to PeV energies is so far lacking, but X-ray synchrotron and TeV $\gamma$-ray observations have revealed super-TeV particles in shocks of young SNRs, e.g. \citet{Ah2004}. If the TeV spectra trace the dominant hadronic CR particles, they should thus extend unbroken to the highest accessible $\gamma$-ray energies if particles with PeV energies are indeed present. The fact that all well-measured TeV spectra of young SNRs show however relatively soft spectra (photon index $\Gamma > 2$) or cutoffs towards higher energies needs to be interpreted through one of two possible scenarios. Either, the spectra are dominated by Inverse Compton (IC) emission from the TeV electrons that are seen in synchrotron X-rays and thus suffer from synchrotron cooling; then, the cutoff energy of the hadronic CRs in these SNR would experimentally be unconstrained, unless a second component above the measured leptonic cutoff energy would be detected in future. Or, the known TeV-emitting shells do no longer confine previously accelerated PeV (hadronic) particles, those would have escaped at earlier times of the SNR evolution and are at present already diffusing into the SNR's environment. The latter hypothesis has been discussed in several works as a possibility to lower the high energy cutoff of TeV spectra of young SNRs in hadronically-dominated TeV emission scenarios \citep[e.g.][]{BerezhkoCasA,Malkov2005,BerezhkoVelaJr}. 

In general, previously accelerated CR particles are assumed to be trapped inside the SNR until the shock significantly slows down and cannot confine the CRs with a given threshold energy any more. Such energy-dependent but purely diffusive escape is possibly at present time the dominant escape process for TeV and super-TeV CRs at middle-aged SNRs such as W\,28, which are not any more accelerating TeV particles. For young, TeV-emitting SNRs (like RX\,J1713.7-3946 and Vela Jr.), direct CR escape from the acceleration region \citep{Zirakashvili2005,Gabici2007} is likely the dominant escape process. This offers an experimental access to identifying these particles as hadrons.
At and outside of the forward shock locations, the TeV morphology of these SNRs might differ from the X-ray morphology (which is dominated by electron synchrotron emission), namely they might extend further outwards. Such effect might be strongly enhanced if the SNR is embedded in areas of dense gas like molecular clouds, which serves as target material for $\pi^{0}$-production preceding the $\gamma$-ray emission \citep{ Ah1996}. Besides such a morphological signature, also a spectral signature could be expected from TeV emission regions outside SNRs, 
since the energy of the escaping particles is depending on the SNR evolutionary stage including the current forward shock speed \citep[e.g.][]{Zirakashvili2005,Gabici2007,Gabici2009};
in essence, the spectra may peak at higher energies further away from the SNR.

In the work presented in this paper, we focus on the TeV-emitting SNR HESS\,J1731-347 \citep{Ah2008b,Ab2011}. This SNR is very similar to the well-known TeV SNRs RX\,J1713.7-3946 and Vela Jr., regarding their physical size and TeV luminosity, in displaying low surface-brightness radio emission \citep{Tian2008}, and by exhibiting essentially pure non-thermal X-ray emission \citep{Ab2011,suzaku2012}. The high forward shock speed ($\gtrsim 1000\,\mathrm{km\,s^{-1}}$) inferred from the non-thermal X-ray emission indicates that the remnant is still in an early evolutionary stage, with super-TeV particle acceleration ongoing in the shocks. A distinctive feature of the HESS\,J1731-347 environment is another resolved TeV source, HESS\,J1729-345, located just outside of the SNR \citep{Ab2011}. This source is in apparent spatial coincidence -- at least in projection -- with molecular clouds seen through sub-mm molecular line emission \citep{Ab2011}. Since no other local particle accelerators are known so far, a scenario is conceivable in which particles that have escaped the SNR HESS\,J1731-347 are presently penetrating the molecular clouds coincident with HESS\,J1729-345 and thus lead to enhanced $\gamma$-ray emission, above the emission induced by the sea of cosmic rays that are homogeneously filling the Galaxy. 

In fact, a similar scenario has been successfully invoked to explain the $\gamma$-ray emission region HESS\,J1800-240 south of the SNR W\,28 \citep{Ah2008,Fermi2010,Li2010}. HESS\,J1800-240 displays a very good morphological match to molecular gas measured through CO emission from that region. 
In contrast to W\,28, HESS\,J1731-347 is however much younger and is presently still accelerating super-TeV particles. It offers the opportunity to model the evolution of the SNR and its associated particle acceleration history, simulating the particle acceleration and escape history up to the present time with good precision. For this purpose, we adopt the particle acceleration and escape formalism developed by \citet{Zirakashvili2008b} and couple it to simulated SNR evolution histories. 

The aim of the study presented in this paper was to explore if the observational constraints are compatible with a scenario in which CRs that have been accelerated in HESS\,J1731-347 and are now diffusing outwards could indeed explain the $\gamma$-ray emission seen from HESS\,J1729-345. To this extent, SNR evolution models were explored (Sect.\,\ref{Subsect:Progenitor} and \ref{Subsect:Evolution}) that are consistent with the known properties of HESS\,J1731-347 (Sect.\,\ref{Subsect:1731}). Using these models, CR acceleration in the forward shock and the escape of the highest energy particles was simulated, adopting the receipes developed by \citet{Zirakashvili2008b} (Sect.\,\ref{Subsect:Evolution}). Over the history of the SNR, the diffusion of the escaping CR particles into an environmental molecular cloud setup was simulated, that was constructed to match the known observational constraints (Sect.\,\ref{Subsect:CRsoutside}). In Sect.\,\ref{Sect:Results}, we discuss two possible scenarios: one in which the SNR still resides in the main sequence wind bubble of the progenitor star (Sect.\,\ref{Subsect:InsideBubble}), and one in which the SNR has started to enter the surrounding wind bubble shell (Sect.\,\ref{Subsect:InsideShell}).

In our model, once the particles have left the SNR, they propagate diffusively and isotropically into the surrounding simulated molecular cloud environment. Such diffusion has been the basis of most corresponding work in the literature \citep[e.g.][]{ Ah1996,Gabici2007,Li2010}. CR particles can normally only be assumed to travel diffusively on sufficiently large scales, $\gg 100\,\mathrm{pc}$ according to the observed maximum scale of magnetic fluctuations in our galaxy \citep[e.g.][]{Han2004}. In general, previous works found that the diffusion speed required for a self-consistent picture requires low diffusion coefficients, $\sim 10$ times lower than typical Galactic values \citep[e.g.][]{Gabici2007,Ah2008,Gabici2009,Li2010,Li2012}, justifying to some extent the concept of diffusive CR transport even on scales $\lesssim 100\,\mathrm{pc}$. For our model of the HESS\,J1731-347 / HESS\,J1729-345 system, resulting diffusion speeds turn out to be closer to typical Galactic values. This is somewhat at odds with the presented geometric solution that places the molecular cloud corresponding to HESS\,J1729-345 at only 30\,pc distance to the SNR. We discuss the implications of this apparent tension in Sect.\,\ref{Subsect:Diffusion}.
The main conclusions of the paper are summarized in Sect.\,\ref{Sect:DiscussionConclusion}.

\section{Model}
\label{Sect:Model}

\subsection{HESS\,J1731-347 and HESS\,J1729-345: observational constraints}
\label{Subsect:1731}

Shell-type emission from the SNR HESS\,J1731-347 is seen through TeV $\gamma$-ray emission \citep{Ah2008b,Ab2011}, as well as through radio synchrotron \citep{Tian2008} and X-ray synchrotron \citep{Ab2011,suzaku2012} emission. In the GeV-(Fermi-LAT) band, only upper limits were derived \citep{Yang2014,Fermi2015}; also, no thermal X-rays have been detected yet from the SNR \citep{Ab2011,suzaku2012}. A morphological match of an absorption pattern in X-rays with a gas density pattern measured through $\rm{^{12}CO}$ column density from the CfA survey \citep{Dame2001} permitted to set a lower limit on the distance to the SNR of $\sim$$3.2\,\mathrm{kpc}$, likely locations are thus in the Scutum-Crux Galactic spiral arm ($\sim$$3\,\mathrm{kpc}$) or the Norma-Cygnus arm ($\sim$$4.5\,\mathrm{kpc}$) \citep{Ab2011}. A luminous point source has been identified in thermal X-rays at the geometrical center of the SNR and has been classified as the associated central compact object (CCO), i.e. a cooling neutron star (NS) remaining after the SN explosion \citep{Ab2011,Klochkov2013}. Modelling of the NS atmosphere indicates that the nearby distance is preferred \citep{Klochkov2013,Klochkov2015}. An association with the nearby H\,II region at $3.2 \pm 0.8\,\mathrm{kpc}$ distance \citep{Tian2008} is thus possible. 

Throughout this paper, we adopt the distance estimate of $3.2\,\mathrm{kpc}$ from Earth, placing the SNR just behind the molecular cloud structure at $3.2\,\mathrm{kpc}$ at the rear side of the Scutum-Crux arm. The radius of the SNR is thus about $15\,\mathrm{pc}$. We note that in a recent work by \citet{Fukuda2014}, a distance of $\sim$$5.2\,\rm{kpc}$ was derived  by matching the \rm{TeV} image with the interstellar gas density profile (using $\rm{^{12}CO}$ data from Nanten and H$\rm{_I}$ data). We discuss this possibility a bit further in the context of our model where the shock has entered the progenitor wind bubble shell (Sect.\,\ref{Subsect:InsideShell}).

The SNR outer shock speed is not known from direct observations. However, the detection of synchrotron X-rays imply forward shock speeds $\gtrsim 1000\,\mathrm{km\,s^{-1}}$, using the argument that the synchrotron peak energy does not depend on magnetic field strength but only on shock speed \citep[e.g.][]{Ah1999,Vink2013}. This theory is also supported by the observation that measured shock speeds of X-ray synchrotron emitting SNRs are always high ($\approx 2000-6000\,\mathrm{km\,s^{-1}}$, \citet{Vink2012}). 

The nearby TeV source HESS\,J1729-345 has a significantly lower surface brightness than HESS\,J1731-347, with moderate peak detection significance ($\sim$$8\,\sigma$, \citet{Ab2011}). Thus, we believe a detailed morphological correlation analysis of the $\gamma$-ray source HESS\,J1729-345 with gas density maps is premature. As shown in \citet{Ab2011}, the source coincides in projection with two molecular cloud structures, one at LSR velocities integrated from $-13$ to $-25\,\mathrm{km\,s^{-1}}$ (at a distance of $\sim$$3.2\,\mathrm{kpc}$, i.e. the foreground cloud), and one at $-75$ to $-87\,\mathrm{km\,s^{-1}}$ (located at a distance of $\sim$$6\,\mathrm{kpc}$ \citep{Ab2011}, i.e.\ likely in the 3\,kpc spiral arm \citep[see][]{Fukuda2014}). A similar range of possible distances is also derived by \cite{Maxted2015} through possible associations with high-density gas ($n_{\rm{H}} \gtrsim10^4 \, \rm{cm^{-3}}$) traced by CS(1-0) data.

As we adopt the distance to the SNR of $3.2\,\mathrm{kpc}$ and assume that HESS\,J1729-345 is a $\gamma$-ray source near the SNR in space, we associate HESS\,J1729-345 with the molecular clouds seen in the distance range $-13$ to $-25\,\mathrm{km\,s^{-1}}$.
The moderate resolution of both the gas density and the TeV $\gamma$-ray maps leaves room to set up a geometrical Molecular Cloud model as introduced in Sect.\,\ref{Subsubsect:Cloudsnearby}. This model is chosen such that it matches the TeV data that are computed in the framework of our CR propagation model, while at the same time not violating any constraints from the gas density data.

\subsection{The progenitor star of a core-collapse SNR and its environment}
\label{Subsect:Progenitor}

The presence of a CCO suggests a core collapse supernova (SN) scenario for HESS J1731-347, implying a progenitor star with relatively large mass, $M>8\,\rm{M_\odot}$.
Massive stars are normally formed as clusters inside MCs which can be considered as MC clumps floating inside the inter-clump medium (ICM): 
clumps have densities around $10^3\sim 10^4\,\rm{H\,cm^{-3}}$, but only take $2\%-8\%$ of the volume; the pressure in the ICM is around $10^5\,\rm{Kcm^{-3}}$ and the density is around $5-25 \,\rm{H\,cm^{-3}}$ \citep{Chevalier1999}.  Thus the ICM is the environment into which the stellar wind bubble and the SNR actually expand. 
Before the SN exploded, the massive progenitor star during its main sequence (MS) phase has likely blown a huge bubble (with a size of up to tens of parsec) in the MC or the ICM. 
After the MS phase,  the massive star will enter the Red Super-Giant (RSG) phase.
For SNe IIP, the RSG wind extends only to $\lesssim1\,\rm{pc}$, while
SNe IIL/b will have larger RSG wind bubbles above $5\,\rm{pc}$ with material up to the CNO layer striped up from the star and extending into the bubble. After the RSG phase, progenitor stars of SNe Ib/c are expected to go through a Wolf-Rayet(WR) phase, during which the high speed WR wind could blow away the dense RSG bubble and form a very turbulent structure inside the MS bubble \citep{Chevalier2005}.

Thus right before the SN happens, we divide the progenitor environment of HESS J1731-347 into several regions: 

\begin{itemize}

\item $ICM$. 
We adopt $n_{\rm{ICM}}=5 \,\rm{H\,cm^{-3}}$ for the ICM density and the standard Galactic diffusion coefficient ($D(E) = D_{10}(E/10\,\rm{GeV})^{\delta}, $ $D_{10}=10^{28}\,\rm{cm^2/s},\ \delta= 0.3\sim0.5$) in the ICM 

\item $MC\ clumps$. 
High gas densities are expected inside MC clumps which can be deduced from the CO data. Diffusion coefficients in the MC clumps are unknown, we will discuss this issue further below.

\item $The\ MS\ wind\ bubble$. 
Starting close to the star, the density in the MS bubble first follows a power-law in the stellar wind region; then it remains constant inside the long shocked wind region ($n_\mathrm{b} \approx 0.01 \,\rm{H\,cm^{-3}}$); at the border of the MS bubble, ICM swept-up by the MS wind forms a MS bubble shell \citep{Weaver1977}. In our models, we ignore the power-law wind region due to its small size and low density compared to the whole MS bubble. 
The bubble shell is assumed to have a thickness $\sim1\,\rm{pc}$. Turbulence is strong in this shocked wind region, thus lower diffusion coefficients than inside the ICM region may be assumed in the following chapters. The magnetic field is assumed to be $B_0=5\,\mathrm{\mu G}$ for CR acceleration calculations, consistent with \citet{Berezhko2000} and \citet{Zirakashvili2005}.

\item $The\ RSG\ wind\ bubble$. 
Compared to the MS wind bubble, the size of the RSG bubble is smaller but its density is much higher. The density follows $n_\mathrm{RSG}(r)= \dot{M}_\mathrm{RSG}/4\pi r^2 v_\mathrm{RSG}$ \citep{Chevalier2005}. 
Here, $\dot{M}_\mathrm{RSG}$ and $v_\mathrm{RSG}$ are the mass loss rate and wind speed during RSG phase.
The bubble shell between the RSG wind bubble and the MS bubble is ignored in our calculations because it only contains very little mass swept up from the almost empty MS bubble.
The turbulence in the RSG wind is also strong. The magnetic field is assumed to be $B_0(r)=2\times 10^{13} (v_\mathrm{RSG}/10^6cm/s)^{-1}(r/cm)^{-1}\,\rm{G}$ \citep{Zirakashvili2005} which is similar to the interplanetary magnetic field \citep{Parker1958}.

\item $The\ WR\ wind$. 
We also explored SNe Ib/c scenarios which imply the presence of a WR wind phase in the progenitor environment. However, we did not explore the complex inhomogeneous structures that are typical for the instabilities evolving from WR winds expanding into RSG wind bubbles. The WR wind together with the RSG wind can only inject very little material into the MS wind bubble. E.g.\ before the SN explosion, a $25\,\rm{M}_\odot$ star will blow about $21\,\rm{M}_\odot$ material into the MS wind bubble (with a size of $\sim21$\,pc) and leaves a CNO core behind. When the WR wind-driven material collides with the MS bubble shell, and if we assume that all material bounces back into the MS bubble, the maximum density contribution after dissipation in the MS bubble is about $0.02\,\rm{H\,cm^{-3}}$. Thus, we adopt a homogeneous but very tenuous density for the WR wind, extending up to the inner boundary of the MS bubble.
The magnetic field is assumed to be the same as in the MS bubble.
\end{itemize}

\subsection{The SNR evolution history and particle acceleration process}
\label{Subsect:Evolution}

With the adopted distance of HESS J1731-347 from Earth of 3.2\,kpc, the radius (defined by the forward shock) of the SNR is about 15\,pc at present.
After starting the expansion, e.g. for type IIL/b SN, the SNR forward shock first passes the relatively dense RSG bubble, then enters the MS bubble, and will finally encounter the ICM.  

For the ejecta-dominated phase of an SNR, a self similar solution was found by \citet{Chevalier1982} and \citet{Nadezhin1985}, with $R_\mathrm{SNR} \propto t^{4/7}$ for an SNR expanding in a uniform medium and $R_\mathrm{SNR} \propto t^{7/8}$ for an SNR expanding in a RSG wind. Here we adopt the derived equations which follow these power-law rules from \citet{Zirakashvili2005}. 
Once the shock leaves the RSG wind and enters the essentially empty MS bubble, if it is at this stage still in ejecta-dominated phase, we simply assume that the shock maintains its speed until it enters the Sedov phase (swept-up mass equals the ejecta mass, $M_\mathrm{s}\approx M_\mathrm{ej}$). To smoothen the transition between two SNR phases, the swept-up mass at this transition phase 
is adapted slightly around $1M_{\mathrm{ej}}$, see also Appendix B.

At Sedov phase, to solve the SNR evolution in an inhomogeneous but spherically symmetric circumstellar medium,  a ``thin-shell'' approximation of a SNR evolution was derived in \citet{Ostriker1988}, \citet{Bisnovatyi1995}. From momentum conservation and ignoring the pressure in the circumstellar medium, \citet{Zirakashvili2005} have derived:
\[
v_{\mathrm{SNR}}(R_{\mathrm{SNR}})=\frac{\gamma_{\mathrm{ad}}+1}{2}\left[
\frac{12(\gamma_{\mathrm{ad}}-1)E_{\mathrm{ej}}}{(\gamma_{\mathrm{ad}}
+1)M^{2}(R_{\mathrm{SNR}})R_{\mathrm{SNR}}^{6(\gamma_{\mathrm{ad}}
-1)/(\gamma_{\mathrm{ad}}+1)}} \cdot \right.
\]
\[
\left. \cdot \int_{0}^{R_{\mathrm{SNR}}}drr^{6\left(
\frac{\gamma_{\mathrm{ad}}-1}{\gamma_{\mathrm{ad}}+1}\right)  -1}M(r)\right]
^{1/2},\; 
\]
\[
t(R_{\mathrm{SNR}})=\int_{0}^{R_{\mathrm{SNR}}}\frac{dr}{v_{\mathrm{SNR}}(r)},
\]
where $\gamma_{\rm{ad}}=4/3$ is the ratio of the
specific heats (adiabatic index), 
$M=M_{\rm{ej}}+4\pi\int_{0}^{R_{\rm{SNR}}}drr^{2}\rho(r)$
is the total mass confined by the shock of radius $R_{\rm{SNR}}$,
$M_{\rm{ej}}$ is the ejected mass, and $\rho$ is the density of the
ambient gas.

With the shock speed $v_{\mathrm{SNR}}$ and the density of the nearby circumstellar medium obtained, 
the CRs acceleration processes can be calculated through the non-resonant acceleration theory from \citet{Zirakashvili2008b}. 
Inside the upstream acceleration region with thickness about $5-10\%\,R_\mathrm{SNR}$ \citep{Zirakashvili2012}, 
the magnetic turbulence is amplified by accelerated CRs, which finally
increases the escape energy $E_\mathrm{max}$. 
We adopt the concept of an absorption boundary with position $L = t_\mathrm{SNR} v_\mathrm{SNR}$ from \citet{Zirakashvili2008b}. $L$ is measured from the shock position $R_\mathrm{SNR}$ in radial direction outwards and is set to the scale where CRs at the high-energy cut-off $E_\mathrm{max}$ cross the boundary and escape into the surrounding medium with a flux $J(E)$ (the number of particles crossing the absorption boundary per unit time and unit surface area). $L$ is of similar order as $R_\mathrm{SNR}$. 

From the acceleration region 
$R \sim 1.1\,R_{\mathrm{SNR}}$ to the absorption boundary $R = R_{\mathrm{Abs}} = R_{\mathrm{SNR}} + L$ is the region where the escaping CRs are driving streaming instabilities before the shock is arriving.
\citet{Zirakashvili2008b} provide an analytical approximation for calculating $E_\mathrm{max}$ along with the CR density at the shock front $N_0(E)$ and the escaping CR flux $J(E)$, where $N_0(E)$ follows a power law with index $\Gamma=2$ in Sect.\,\ref{Subsect:InsideBubble}. We adopt these prescriptions for the work presented here. More details about the acceleration model can be found in Appendix A.

\subsection{Cosmic Rays outside of the SNR}
\label{Subsect:CRsoutside}

\subsubsection{Analytical diffusion model}
\label{Subsubsect:Analyticaldiffusion}
CRs escape the SNR starting from the absorption boundary which is assumed to be a spherically symmetric surface surrounding the SNR. In a homogeneous environment, 
the final CR density at distance $R$ from the SNR center
and SNR age $t_{\rm{age}}$ can simply be written as:
\[
n_{\mathrm{integ}}(E,R,t_\mathrm{age}) = \frac{1}{2}\int^{t_{\mathrm{age}}}_{0} dt\int^{\pi}_{0}  sin\theta   J_{\mathrm{total}}(E,t) G(E,R',t_{\mathrm{age}}-t ) d\theta \ ,
\]
where $G(E,r,t)=1/8(\pi t D)^{-3/2} \exp[-r^2/(4 tD)] \ \ ( \int_0^\infty dr 4\pi r^2  G=1)$ is the Green's function solution for the diffusion equation in homogeneous environments ($D=$ const.), see e.g.\ \citet{Ah1996}. $R'=\sqrt{[R-(R_{\rm{Abs}}-R_{\rm{Abs}}\cos\theta)]^2 + (R_{\rm{Abs}}\sin\theta)^2}$ is the distance from a certain point on the absorption boundary to the diffusion target 
(e.g. the nearby molecular clouds),
 $R_{\rm{Abs}}=R_{\rm{SNR}}+L$ is the radius of the absorption boundary, and $J_{\rm{total}}= 4\pi R_{\rm{SNR}}^2 J$ \footnote{Here we use $R_{\rm{SNR}}$ instead of $R_{\rm{Abs}}$, because $J$ is calculated through a plane-parallel shock approximation.} 
is the escaping CR flux at different times along the SNR evolution history.
\subsubsection{Monte-Carlo diffusion model}
\label{Subsubsect:MCdiffusion}

The diffusion coefficient is usually assumed to be a power-law function of energy,
$D(E)=D_{10}(E/10\,\rm{GeV})^{\delta}$. Under realistic conditions, both $D_{10}$ and $\delta$ depend on the environment, e.g.\ in the MS wind bubble or in irregularly shaped MCs. Thus, we use Monte-Carlo simulations to divide the particle diffusion process into small steps. During a short period $\Delta t$, CRs can be considered to diffuse inside a homogeneous environment. For each step, particles are assumed to be scattered by turbulent magnetic fields to a random direction, along this direction they travel the mean diffusion distance $\Delta \bar{r} = 2.5 \sqrt{\Delta t D}$. 
The coefficient 2.5 has been derived by matching results from the Monte-Carlo method with the corresponding analytical expressions for a homogeneous environment; a similar result ($\Delta \bar{r}= \int_0^\infty dr \cdot 4\pi r^2 \cdot r \cdot G \approx 2.3 \sqrt{\Delta t D}$) is obtained when integrating the Green's function.

Each step should be much smaller than the size of the molecular clouds used in the simulated environment, $\Delta r << R_\mathrm{MC}$. In our work, the step size is set to $\Delta r = 1\,\rm{pc}$.

For the CR particle diffusion, the environment is for simplicity divided into three domains with different diffusion coefficients: the MS bubble (including the RSG or eventually the WR wind region as well as the temporally-expanding SNR region, all set to have the same diffusion coefficient), the MC clumps region, and the ICM region.  
An impact of the different diffusion coefficients on the shock acceleration efficiency, e.g. through returning CRs, is estimated to be minor and is ignored in the simulations.

\subsubsection{The nearby Molecular Clouds}
\label{Subsubsect:Cloudsnearby}
\begin{figure}[h!]
\centering
\begin{tabular}{c}
\hspace{-.5cm}
{\includegraphics[width=7.7cm]{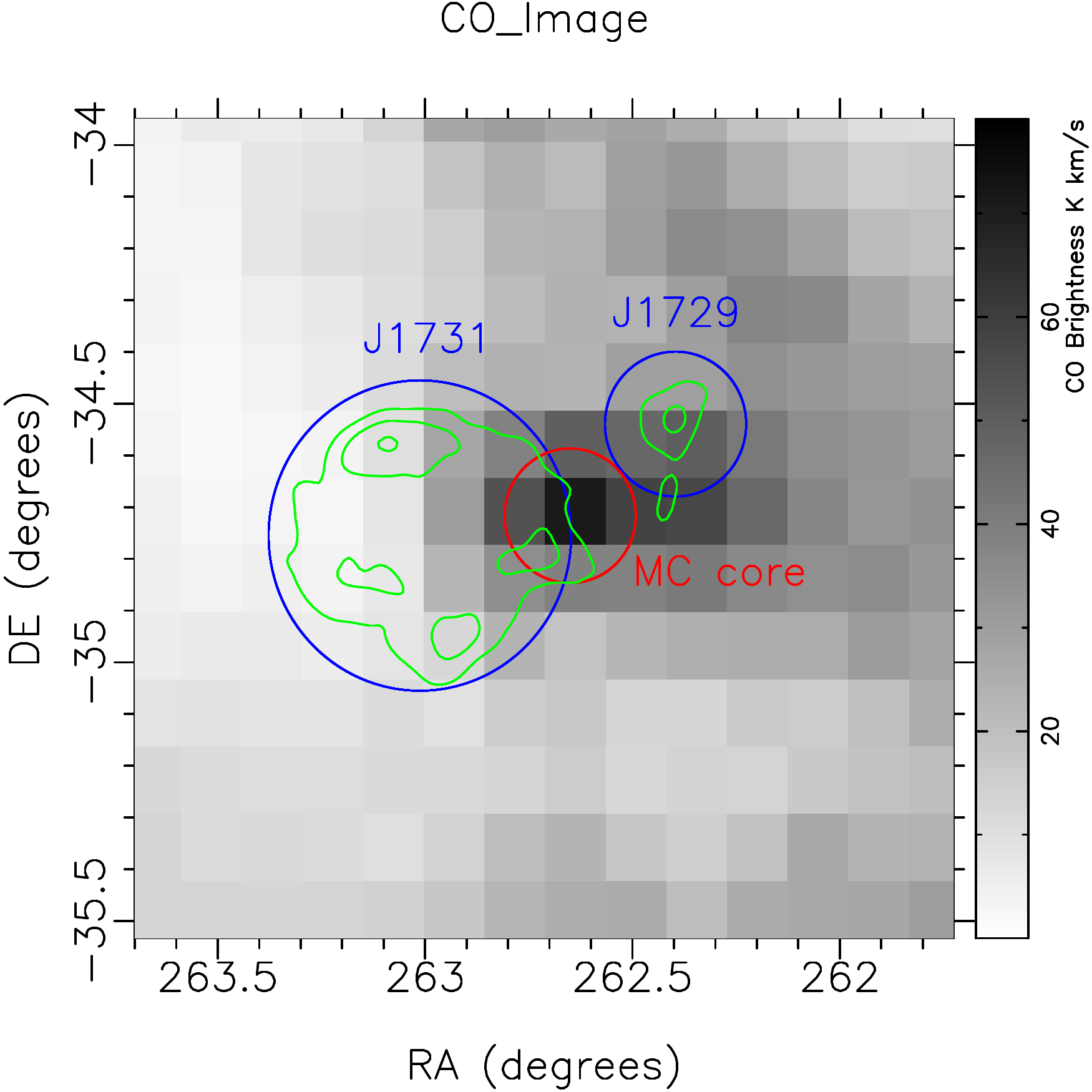} }  \\
{\includegraphics[width=4.1cm]{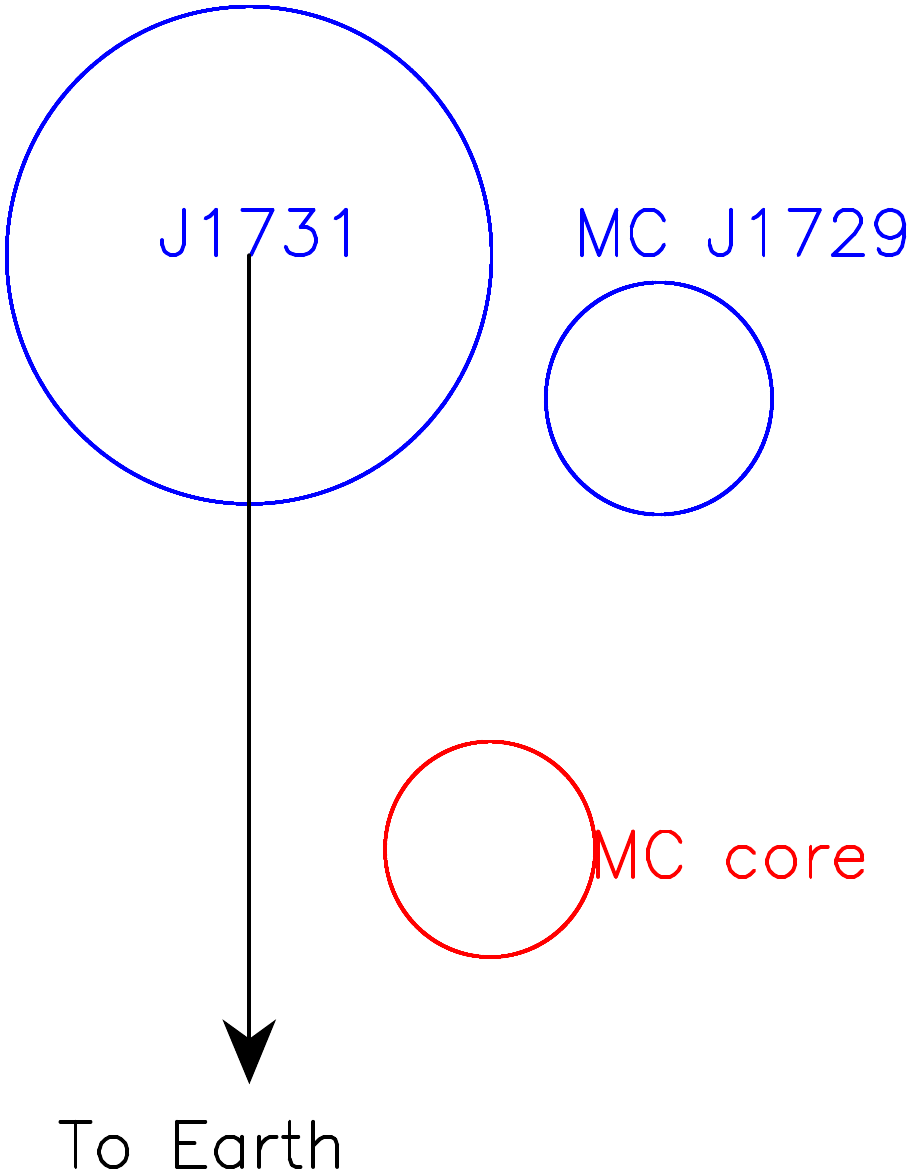} }  
\end{tabular}
\caption{
The top image shows in greyscale the molecular gas density along the line of sight to HESS J1731-347 in a distance interval 2.5\,kpc- 4.0\,kpc, derived by integrating the $\rm{^{12}CO}$ emission from the CfA CO survey from $-13\,\rm{km/s}$ to $-25\,\rm{km/s}$ \citep[same interval as used in][]{Ab2011}.
The big and small blue circles represent the regions from where the observational spectra of HESS J1731-347 and HESS J1729-345 were obtained in \citet{Ab2011}, respectively.
The red circle represents the position of the densest core of the modelled molecular cloud. 
The bottom sketch depicts the modelled geometry with a view from above the galactic plane.
}
\label{fig:CO}
\end{figure}

As discussed in the introduction, the goal of this work is to explore if a scenario is feasible in which the TeV-emitting region HESS J1729-345 can be explained by CRs that escape the young SNR HESS J1731-347 and are currently diffusing away from the SNR. To that extent, we therefore constructed a simplified molecular cloud setup that is consistent with the current molecular line observations and can explain the boost of TeV emission towards HESS J1729-345 through $\pi^0$-decay of CRs interacting with the dense gas. 

The gas of the molecular clouds shown in Fig.~\ref{fig:CO} is likely concentrated in relatively dense clumps and filaments with sizes $\lesssim10\,\rm{pc}$. The most simplified representation of the overall molecular cloud structure that satisfies our goal consists of two spherical, homogeneously filled clumps, one corresponding to the HESS J1729-345 region (MC-J1729, radius 7.8\,pc, see small blue circle in Fig.~\ref{fig:CO}) and another one corresponding to the densest molecular cloud core region (MC-core, centered at R.A. 17h30m36s, Dec -34$^\circ$43'0" and with an ad-hoc set radius of 7.2\,pc, see red circle in Fig.~\ref{fig:CO}). 
To translate the measured CO intensity of these two clumps into corresponding gas density, we adopt the CO-to-H$_2$ mass conversion factor $1.8\times10^{20}\,\rm{cm^{-2}K^{-1}km^{-1}s}$ \citep{Dame2001}, and obtain mean densities of $367 \,\rm{H_2\,cm^{-3}}$ (MC-core, total mass $2.84\times 10^4\, \mathrm{M_{\odot}}$) and $240\,\rm{H_2\,cm^{-3}}$ (MC-J1729, $2.36\times 10^4\, \mathrm{M_{\odot}}$).

As can be seen in the top panel of Fig.~\ref{fig:CO}, the gas density and the \rm{TeV} observational image do not match too well. In our setup, this is a consequence of the non-uniform density of the CRs injected by the SNR, following the injection and diffusion history over the lifetime of the SNR. Putting MC-J1729 closer in space to the SNR than the bulk of the gas material (represented in our setup by MC-core) will result in a TeV emissivity dominated by MC-J1729. MC-J1729 is set to be at the minimum possible distance, i.e. the projected 30.7\,pc distance to the SNR, while MC-core is placed at a distance of 100\,pc in the foreground of the SNR. 

In our work, we simplify the hadron collisions between CRs and target atoms (e.g. in the MC) to proton-proton (pp) collisions only. The cross section table of the pp collision was obtained from \citet{Kachelriess2012}.

\section{Results}
\label{Sect:Results}

\subsection{HESS\,J1731-347 still residing inside the main sequence bubble}
\label{Subsect:InsideBubble}
\begin{figure}
\begin{tabular}{c}
{\includegraphics[width=8.3cm, height=5.8cm]{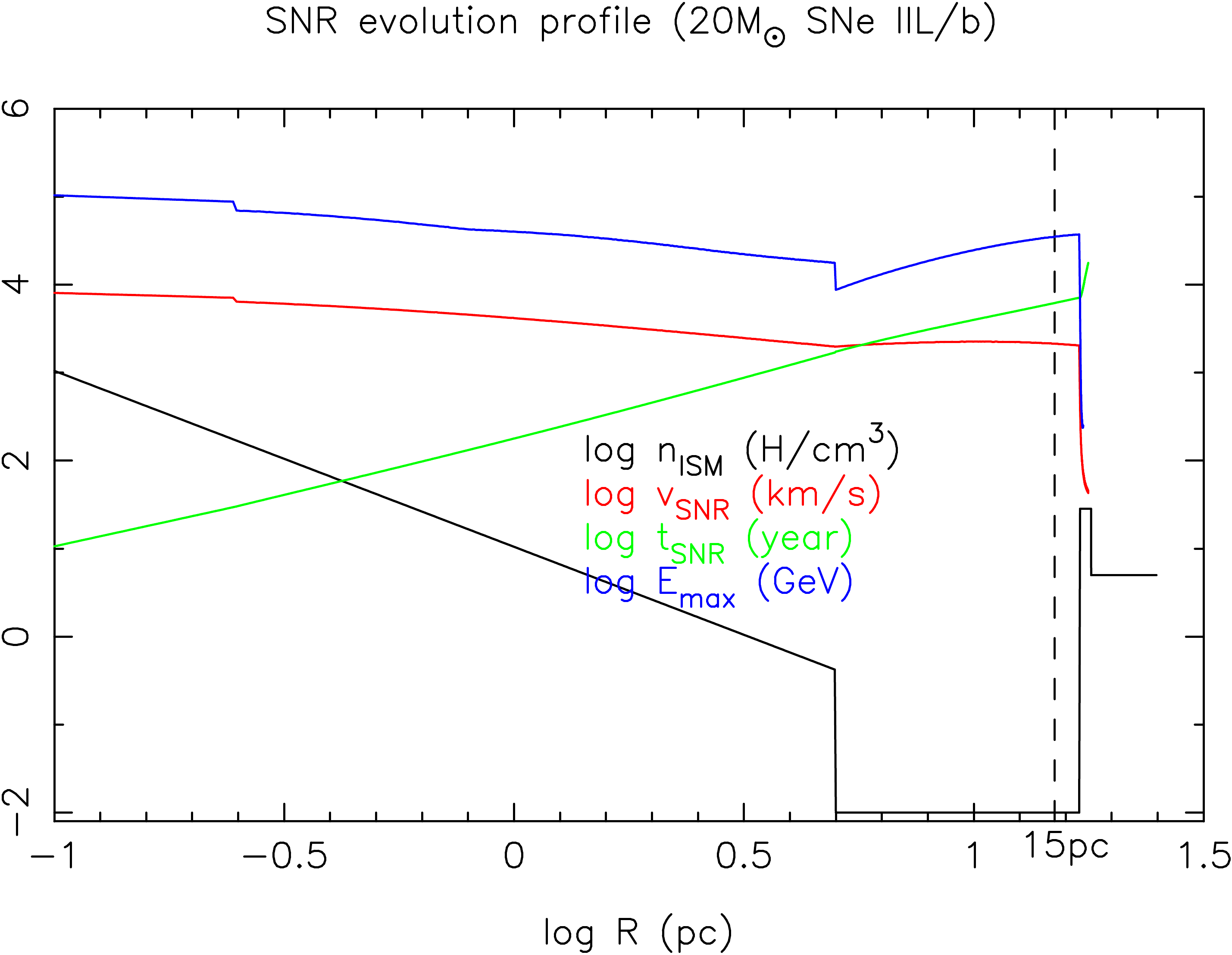} }
\end{tabular}
\caption{The SNR evolution profiles of the SNe IIL/b scenario: density of the circumstellar medium (black), shock velocity (red), age of the SNR (green) and escape energy (blue) along with SNR radius. 
The SNR size (15\,pc) is marked as a dashed line. The time, velocity and escape energy profiles are cut off when shock speed drops significantly ($v_\mathrm{SNR}\ll1000\,\rm{km/s}$). 
}
\label{fig:SNR}
\end{figure}

\begin{figure}[h!]
\centering
\begin{tabular}{c } 
\hspace{-0.5cm}
{\includegraphics[width=7.8cm, height=5.cm]{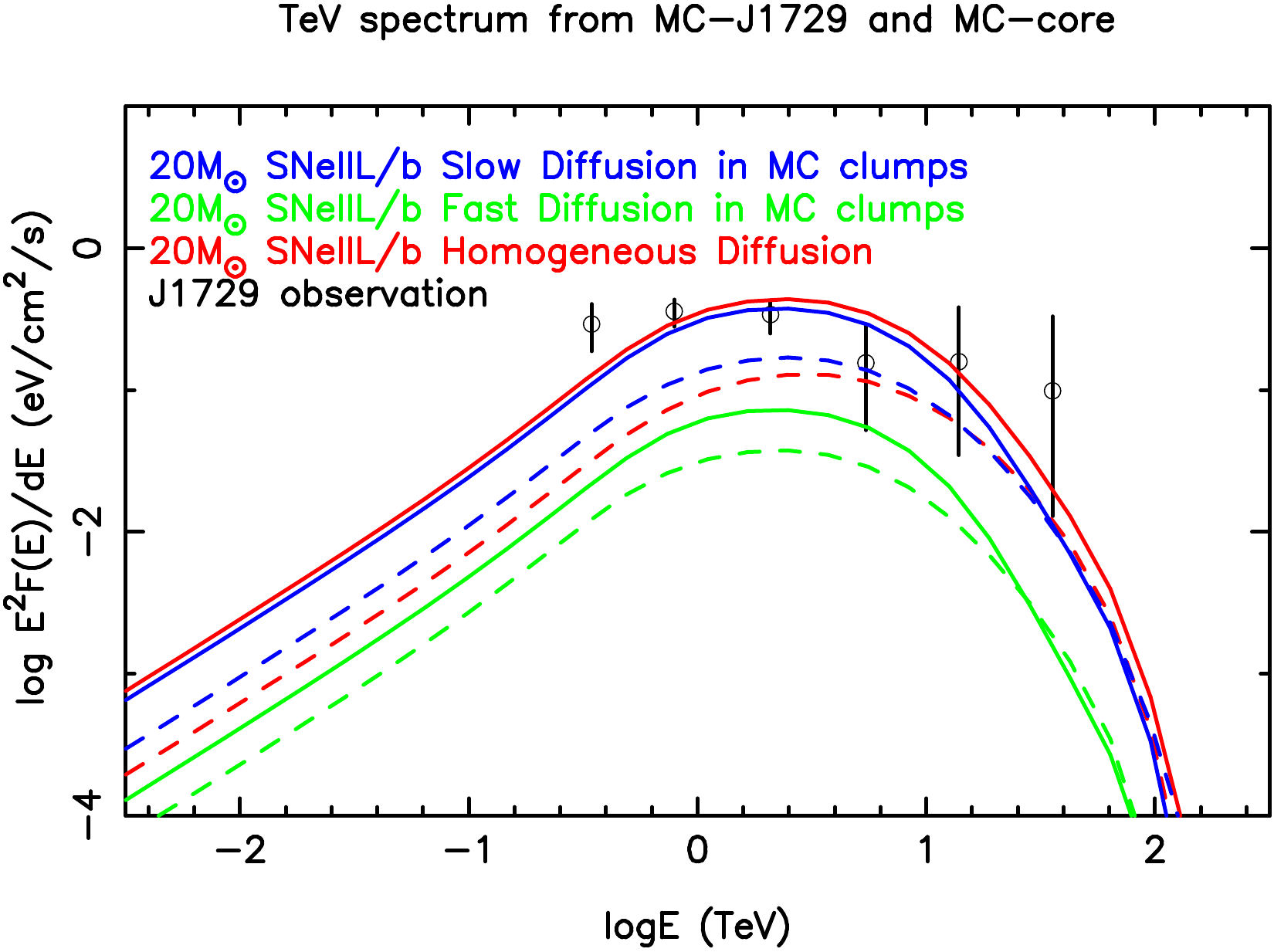} }   \\ \hspace{-0.cm}
{\includegraphics[width=6.2cm]{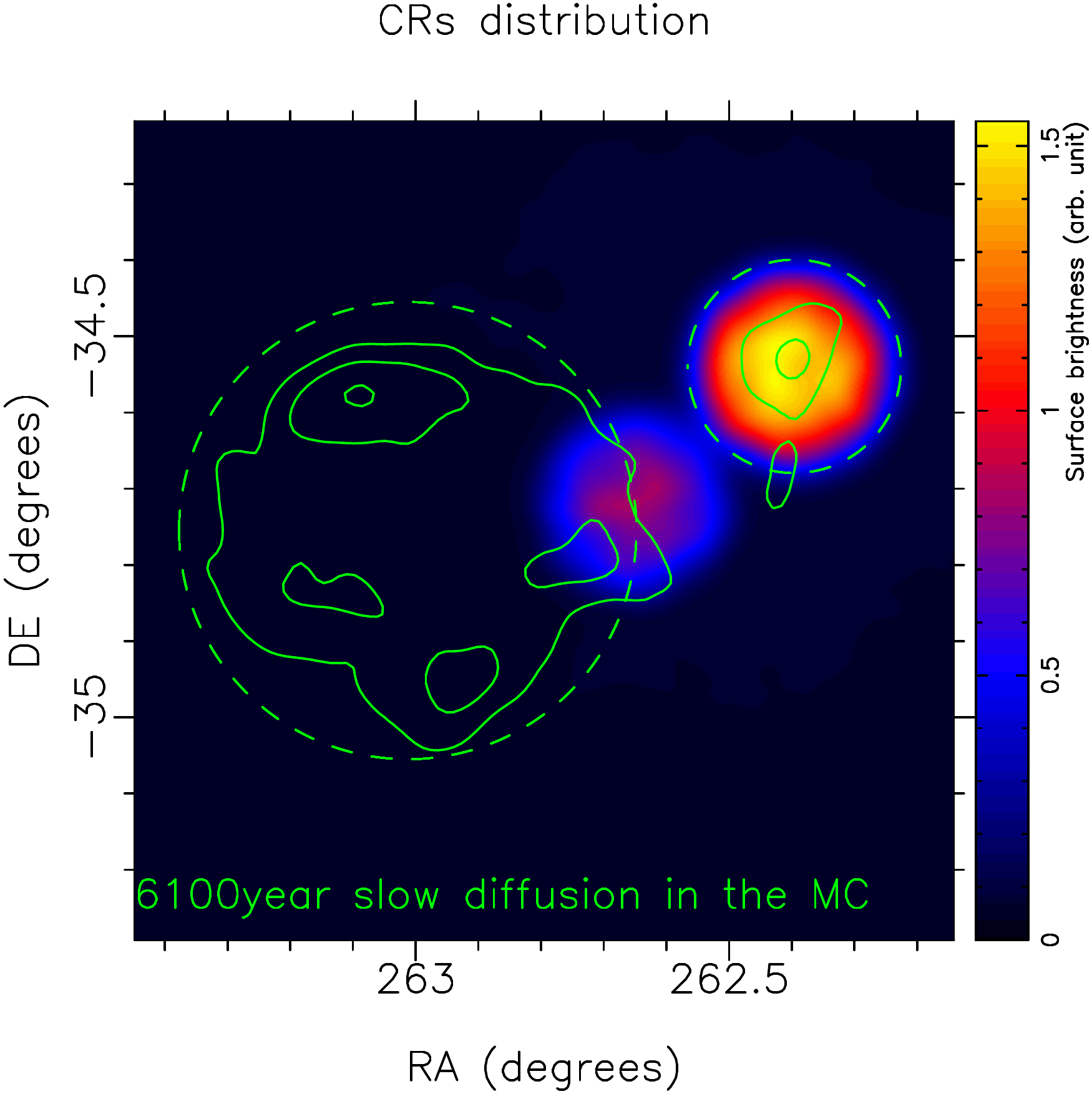} } 
\end{tabular}
\caption{  Predictions for the TeV $\gamma$-ray emission from the escaping CRs in the $20\,\rm{M_\odot}$ progenitor mass scenario,
derived using Monte-Carlo simulations for the CR diffusion into the inhomogeneous medium. On the {\it top panel}, predictions for the TeV spectrum at MC-J1729 (solid lines) and MC-core (dashed lines) are shown, for different diffusion scenarios (red, green, and blue) corresponding to the parameters that are given in Table \ref{table:diffusion}. The {\it bottom panel} sky image shows the predicted $1\,\rm{TeV}$ image for the ``slow diffusion in MC'' scenario (blue lines in the top panel). Green contours correspond to the \rm{TeV} image from \citet{Ab2011}, the big and small dashed circles represent the locations of HESS J1731-345 and HESS J1729-347, respectively.
}
\label{fig:spec}
\end{figure}

In Table 1 we present the parameters for four representative core-collapse SN scenarios (progenitor mass 8$\,\rm{M_\odot}$, 15$\,\rm{M_\odot}$, 20$\,\rm{M_\odot}$, 25$\,\rm{M_\odot}$). The environments for the SNR evolution are adapted as discussed in Sect.\,\ref{Subsect:Progenitor}. 
Under these conditions, the CR escaping and diffusing process is calculated to explain the \rm{TeV} image and spectrum of HESS J1729-345, see Fig.~\ref{fig:SNR} and Fig.~\ref{fig:spec}. In the following, we show detailed results only for the 20$\,\rm{M_\odot}$ scenario. 
The 8$\,\rm{M_\odot}$ and 15$\,\rm{M_\odot}$ turn out to unlikely be realised, mainly because of the currently already low shock speed which is in conflict with the hard X-ray spectrum. 
Under certain simplifications, also a 25$\,\rm{M_\odot}$ scenario can be constructed that satisfies our requirements, but a detailed discussion is beyond the scope of this paper.

\begin{table*}
\caption{SNR evolution of different scenarios}            
\label{table:SNR}     
\centering          
\begin{tabular}{ c c c c c c c c c c c c}     
\hline\hline       
SN Type& $M \ ^{a}$ &
$R_\mathrm{b,MS} \ ^{b}$ & $R_\mathrm{b,RSG} \ ^{c}$ 
&$E_\mathrm{ej}\ ^{d}$& $M_\mathrm{ej}\ ^{e}$& $R_\mathrm{SNR, \ end}\ ^f$ & $t_\mathrm{SNR,\ end}\ ^{g}$ & $v_\mathrm{SNR,\ end}\ ^{h} $& $\eta_\mathrm{esc} \ ^i$ &$E_\mathrm{max,\ end}\ ^j$ &$E_\mathrm{CR, end}\ ^k$\\ 
\hline
SNe IIP& $8 \,\rm{M_{\odot}}$ & 0.5\,pc& - & $1 \,\rm{E_{51}}$ &$6 \,\rm{M_{\odot}}$& 10\,pc & $15.5 \,\rm{kyr}$& $250\,\rm{km/s}$ &0.1 & $6.5\,\rm{TeV}$ & $0.23 \,\rm{E_{51}}$\\
SNe IIP  &$15 \,\rm{M_{\odot}}$ &10\,pc& 1\,pc & $1 \,\rm{E_{51}}$ &$12 \,\rm{M_{\odot}}$& 10\,pc &$5.8 \,\rm{kyr}$& $150\,\rm{km/s}$&0.1 & $0.8\,\rm{TeV}$ & $0.12 \,\rm{E_{51}}$\\
SNe IIL/b  &$20 \,\rm{M_{\odot}}$ &18\,pc& 5\,pc & $1 \,\rm{E_{51}}$ &$2 \,\rm{M_{\odot}}$& 15\,pc &$6.1 \,\rm{kyr}$& $2140\,\rm{km/s}$ &0.02& $34.9\, \rm{TeV}$ & $0.05 \,\rm{E_{51}}$\\
SNe Ib/c  &$25 \,\rm{M_{\odot}}$ &22 pc& - & $1 \,\rm{E_{51}}$ &$2 \,\rm{M_{\odot}}$& 15\,pc &$2.9 \,\rm{kyr}$& $2470\,\rm{km/s}$ &0.01& $16.5\, \rm{TeV}$ & $0.01 \,\rm{E_{51}}$\\
\hline                  
\end{tabular}
\tablefoot{ \\
$^{a}${ Initial mass of the progenitor star.}  \\
$^{b}${ Size of the MS wind bubble (including the MS bubble shell). The numbers were chosen under the reasonable assumption that the pressure of the circumstellar medium is $10^5\,\rm{Kcm^{-3}}$ \citep{Chevalier1999,Chen2013}.}  \\
$^{c}${ Size of RSG wind bubble, corresponding to $\dot{M}_\mathrm{RSG}\approx 0.2(5)\times10^{-5} \,\rm{M_{\odot}}/s$ and $v_\mathrm{RSG}\approx 10(15)\,\rm{km/s}$ for $15\,\rm{M_\odot}$ SNe IIP ($20\,\rm{M_\odot}$ SNe IIL/b) \citep{Chevalier2005}.} \\
$^{d}${ Total SN energy. Core-collapse SNe have observed kinetic en\rm{erg}ies of typically $\sim10^{51} \,\rm{erg}(\,\rm{E_{51}})$\citep{Smartt2009}.}  \\
$^{e}${ Ejecta mass of the SN. The progenitor mass is the sum of the MS wind mass loss, the RSG wind mass loss, the neutron star mass ($2\,\rm{M}_\odot$), and the SN ejecta mass, respectively. }  \\
$^{f}${For the two SNe IIP scenarios, we stop the calculations when the Sedov phase is about to end ($v_\mathrm{SNR}\ll1000\,\rm{km/s}$), the corresponding SNR radius is around 10\,pc. For the SNe IIL/b and Ib/c scenarios, we calculate the SNR history until the forward shock reaches 15\,pc which is the observed radius. } \\
$^{g}${ Age of the SNR when it expands to $R_\mathrm{SNR, \ end}$.}  \\
$^{h}${ Forward shock velocity of the SNR when it expands to $R_\mathrm{SNR, \ end}$. }  \\
$^{i}${ This parameter is the ratio between the energy flux from the escaping CRs and the kinetic energy flux from the upstream medium falling to the shock. This parameter is chosen by fitting the TeV spectrum of HESS J1729-345 as described in section \ref{Sect:Results} and Appendix B.}  \\
$^{j}${ Escape energy of CRs from the SNR shock when the shock radius has reached $R_\mathrm{SNR, \ end}$. } \\
$^{k}${ Total escaped CR flux integrated from the SN explosion to $t_\mathrm{SNR, \ end}$.} \\}
\end{table*}

More than half of all recorded core-collapse SNe are of type IIP, their progenitor star masses range from $8\,\rm{M_\odot}$ to $\sim 25\,\rm{M_\odot}$ \citep{Smartt2009}. In agreement with the lack of high mass RSGs $>15\,\rm{M_\odot}$ exploding as Type IIP SNe (the RSG problem, \citet{Smartt2009}) we choose progenitor stars with masses $8\,\rm{M_\odot}$ and $15\,\rm{M_\odot}$ for our first two scenarios.
In the $8\,\rm{M_\odot}$ scenario, the progenitor star will produce almost no MS wind bubble or RSG wind bubble, thus the SNR will expand directly into the ICM.
In the $15\,\rm{M_\odot}$ scenario, after the SNR forward shock passes the RSG bubble which only contains $\sim0.2\,\rm{M_\odot}$ gas, it continues its ejecta-dominated phase through the whole MS-wind bubble without speed loss. Once the SNR sweeps into the shell of the MS wind bubble, which contains $\sim500\,\rm{M_\odot}$ gas, the shock speed drops significantly.
Both SNe IIP scenarios (with $8$ and $15\,\rm{M_\odot}$, respectively) fail to maintain a high shock velocity until today. More details about these two scenarios can be found in Appendix B. 

Indeed, the MS bubble size is the key parameter to maintain a high shock speed at present time. As long as the SNR is still expanding inside its progenitor MS bubble, the swept-up material is limited and a fast shock speed $v_\mathrm{SNR}>1000\,\rm{km/s}$ can easily be obtained. The bubble size of the progenitor of HESS\,J1731-347 is unknown, as it is for most OB stars in our galaxy due to absorption. Therefore, we adopt the linear relationship between progenitor mass and MS bubble size from \citet{Chen2013}. Their work is based on 15 well-observed OB stars ($8-72\,\mathrm{M_\odot}$) and yields $p_5^{1/3}R_\mathrm{b}=\left[\alpha\left(M/\mathrm{M_\odot} \right)-\beta\right] \rm{pc},$ where $\alpha=1.22\pm0.05$ and $\beta=9.16\pm1.77$. This relationship is also consistent with the theory that a maximum bubble radius $R_\mathrm{b}$ is obtained for pressure equilibrium between bubble and its surroundings, assuming a constant ICM pressure $p/k \approx10^5 \,\rm{cm^{-3}K}$ (i.e., $p_5\approx1$) \citep{Chevalier1999}. In order to confine the observed SNR into a correspondingly larger MS bubble, the progenitor star should have an initial mass of $\gtrsim 20 \,\rm{M_\odot}$.

Stars with progenitor masses above $\sim17\,\rm{M_\odot}$ can end as SNe IIL/b, which represent only $\sim 10\%$ of the total recorded core collapse SNe \citep{Smartt2009}. Here we choose $20\,\rm{M_\odot}$ to make sure the MS-wind bubble ($\sim$17\,pc) just exceeds the observed SNR radius. As shown in Fig.~\ref{fig:SNR}, the SNR enters its Sedov phase in less than 100 years after the SN explosion, when the forward shock is still inside the RSG bubble. With almost no speed loss while moving through the MS-wind bubble, the shock is still a very active acceleration engine at present time ($v_\mathrm{SNR}>2000\,\rm{km/s}$).
 
At the early stage of its evolution, despite the young age of the SNR, the escape energy can easily be as high as $E_\mathrm{max}\sim 100-1000\, \rm{TeV}$, due to the high shock speed and the dense RSG wind the SNR encounters. 
 
A progenitor star with much larger mass ($\gtrsim25\,\rm{M_\odot}$) will develop into a Wolf-Rayet phase after the RSG phase. The ongoing WR phase is characterized by strongly non-uniform, turbulent spatial distribution of the wind material. Indeed, the X-ray morphology of the SNR as shown in \cite{Ab2011} and \cite{suzaku2012} might be indicative of such a phase. 
Since we only model structures in one (radial) dimension, such a turbulent configuration cannot be accommodated for in our model.
Nevertheless, no matter what kind of structure the wind bubble has, the total mass confined inside the MS wind bubble will mainly come from the progenitor star and therefore is very limited. Thus, when the SNR shock is sweeping inside the MS wind bubble, it can also maintain a relatively fast speed just as the SNR in the $20\,\rm{M_\odot}$ scenario does.
Ultimately, the WR wind will blow away the dense RSG wind material into the MS bubble, as discussed in Sect.\,\ref{Subsect:Progenitor}, leaving a tenuous ($n\approx0.02\,\rm{H\,cm}^{-3}$) but roughly homogeneously-filled bubble. Such a configuration can be simulated in our framework (see Tab.\,\ref{table:SNR}), the results confirm the high shock speed and a rough consistency with the TeV spectrum of HESS J1729-347.

As discussed in section 2.4, we use Monte-Carlo simulations with ``regional'' diffusion coefficients that depend on the location where the cosmic rays are moving. We use the Ansatz that the diffusion coefficient $D$ at 10\,GeV corresponds to the average Galactic value ($D_{10} = 10^{28}\,\rm{cm^2/s}$) everywhere, but at higher energies which are in fact relevant for our work here, we introduce differences by choosing different values of $\delta$ in $D(E)=D_{10}(E/10\,\rm{GeV})^\delta$. The values of $\delta=0.3, 0.5$ are still in the range of Galactic diffusion coefficients ($D_{10}=10^{28}\,\rm{cm^2/s}, \delta=0.3\sim0.6$, \cite{Berezinskii1990,Ptuskin2006}). Table 2 lists the values for $\delta$ that have been explored. 

Observationally, the diffusion coefficient in high-density molecular clouds is not well constrained. 
CRs could e.g.\ be ``trapped'' by a lower diffusion coefficient there. For example, \citet{Gabici2010} argued that in order to explain the TeV emission of the molecular clouds near the SNR W\,28, an average diffusion coefficient around 10\% of the galactic standard diffusion coefficient is required, albeit not only in the molecular clouds but everywhere around the SNR. \cite{Crutcher2012} found that the maximum strength of the interstellar magnetic field stays constant at $\sim10\,\rm{\mu G}$ up to densities $n_\mathrm{H}\sim 300\,\rm{cm}^{-3}$, and above $300\,\rm{cm}^{-3}$ it increases following a power-law with exponent $\approx2/3$. If one assumes that the magnetic turbulent strength in MC clumps is of order $\sim10\,\rm{\mu G}$ as well, with a Kolmogorov-type power-law for the magnetic turbulent power spectrum and the size of the MC clumps $\sim$1\,pc as maximum wavelength, one can obtain a much lower diffusion coefficient in dense clump regions, with $\lesssim1\%$ Galactic standard for 1-1000\,TeV CRs \cite[e.g.][]{Fatuzzo2010}.  On the other hand, diffusion in dense molecular clouds could be even much faster than Galactic due to damping of the turbulent magnetic field in the high density environment. The values chosen in our work ($\delta=0.5$ and $\delta=0.3$ as shown in Table 2) correspond to a range of the diffusion coefficient of 1-10\,TeV protons with a mild factor of only $\sim 3$ between minimum and maximum. More extreme values are currently not compatible with our model.

\begin{table}[h!]
\begin{small}
\caption{
\label{table:diffusion}
Diffusion index ($\delta$) in different regions for the Monte-Carlo simulation shown in Fig.~\ref{fig:spec}.}
\begin{tabular}{lcccc}
\hline
 Diffusion(Dif.) regions &MS bubble & ICM & MC clumps  \\ 
 Density  $\rm{H\,cm^{-3}}$     & 0.01  & 5 & 480/734 (J1729/Core) \\
\hline
Homogeneous Dif.& 0.3 & 0.3& 0.3 \\
Fast Dif. in MC clumps& 0.3 & 0.5& 0.5\\
Slow Dif. in MC clumps& 0.3 & 0.5& 0.3\\
\hline

\end{tabular}
\tablefoot{In this work, we only modify the energy index $\delta$ of the diffusion coefficient $D(E)=D_{10}(E/10\,\rm{GeV})^\delta$, while $D_{10}=10^{28}\,\rm{cm^2/s}$ is fixed for all scenarios. 
}
\end{small}
\end{table}

As introduced in section 2.4.3, we place MC-J1729 at the shortest possible distance to the SNR (30.7\,pc), while MC-core is located at the foreground of the SNR, 100\,pc away. Fig.~\ref{fig:spec} shows the simulated TeV spectra for the different assumed diffusion coefficient setups, and a simulated TeV sky map for the ``slow diffusion in MC'' parameters. 
The simulated spectra peak around an average $E_\mathrm{max}\approx 30\,\rm{TeV}$, and a rough match with the measured spectrum of HESS J1729-345 can be achieved with the two models that use the low diffusion coefficient in the MC clumps of $\delta=0.3$.
The simulated sky map illustrates that the expected TeV emission from MC-core is substantially suppressed with respect to the MC-J1729 emission, and thus remains undetectable in the measured TeV sky map due to the dominance of the emission from the SNR itself. This was indeed one main boundary condition for the choice of the model setup parameters in this work. The emission from the SNR itself is likely dominated by Inverse Compton emission from CR electrons, given the low gas density inside the wind bubble in which the forward shock still resides.

Several of the free parameters of our model are obviously correlated:
\begin{itemize} 
\item  The difference between the CR density at MC-core and MC-J1729 can be increased by lowering the ICM diffusion coefficient at the relevant high energies. In turn, if we use a higher ICM diffusion coefficient than the galactic standard value($\delta>0.5$), we would have to put MC-core even further away from the SNR.
\item  With a lower diffusion coefficient in MC clumps like MC-J1729, more CRs will be trapped there, while the CR density at MC-J1729 is not sensitive to the diffusion coefficient in the ICM. For our scenario, the lower diffusion coefficient inside the MC clumps (by choosing $\delta=0.3$ as in ``homogeneous diffusion'' or ``slow diffusion in MC'') matches the measured TeV spectrum reasonably well, see Fig.~\ref{fig:spec}. Nevertheless, also the ``fast diffusion in MC'' setup could be modified to match the TeV data, by increasing either $E_\mathrm{ej}$ or $\eta_\mathrm{esc}$ to generate more CRs from the SNR. 
\item The CRs that have diffused to the MC clumps have a particle energy spectrum that is peaking near the average $E_\mathrm{max}$. To move the peak (and thus also the corresponding TeV $\gamma$-ray peak) to lower energies, the acceleration efficiency can be lowered, or the diffusion coefficient in the ICM can be increased. The distance between MC-J1729 and the SNR cannot be lowered below 30.7\,pc, since this corresponds already to the minimum possible distance.
\end{itemize}

In summary, the model is in general sensitive to the choice of diffusion coefficients in the medium. However, the TeV brightness contrast between MC-J1729 and MC-core (in the foreground of the SNR) can be kept such that it does not violate the TeV data by adjusting several parameters that are not strongly constrained. Therefore, no conclusion on the actual diffusion coefficients in the different media can be drawn in the framework of the presented model.

\subsection{HESS\,J1731-347 evolving into the main sequence bubble shell}
\label{Subsect:InsideShell}

\begin{table*}
\caption{
\label{table:SNR2} 
SNR evolution when SNR evolving into the MS bubble shell.            
}  
\centering          
\begin{tabular}{ c c c c c c c c c c c c c}    
\hline\hline       
Scenarios \ $^{a}$ & $R_\mathrm{b,MS} $ 
&$E_\mathrm{ej}$& $M_\mathrm{ej}$& $R_\mathrm{SNR, \ end}$ & $t_\mathrm{SNR,\ end}$ & $v_\mathrm{SNR,\ end} $&$n_\mathrm{end}\ ^{b} $ & $\eta_\mathrm{esc} $ &$E_\mathrm{max,\ end}$ &$E_\mathrm{CR, end}$ & $E_\mathrm{CR,sh} \ ^{c}$\\ 
\hline
20$\,\rm{M}_\odot$ in-shell  &20.5\,pc & $2 \,\rm{E_{51}}$ &$2 \,\rm{M_{\odot}}$& 15\,pc &$4.9 \,\rm{kyr}$& $1150\,\rm{km/s}$ & $0.58 \,\rm{cm^{-3}}$ &0.02& $15.6 \,\rm{TeV}$ & $0.13 \,\rm{E_{51}}$ & $0.12 \,\rm{E_{51}}$\\
25$\,\rm{M}_\odot$ in-shell   &21.5\,pc& $2 \,\rm{E_{51}}$ &$2 \,\rm{M_{\odot}}$& 15\,pc &$2.4 \,\rm{kyr}$& $1170\,\rm{km/s}$ & $0.35 \,\rm{cm^{-3}}$ &0.02& $6.6 \,\rm{TeV}$ & $0.08 \,\rm{E_{51}}$&  $0.07 \,\rm{E_{51}}$\\
\hline

\hline                  
\end{tabular}
\tablefoot{ \\
$^{a}${ For both scenarios, the same initial progenitor star (scenario $20 \,\rm{M_{\odot}}$, SNe IIL/b and scenario $25 \,\rm{M_{\odot}}$, SNe Ib/c) and the same RSG bubble structure as shown in Table 1 are assumed. They only differ in MS bubble shell structure and SN energy.
} \\
$^b${ The gas density in the upstream of the shock at $t_\mathrm{SNR, \ end}$.
}   \\
$^c${ The total CR energy trapped inside the downstream of the shock at $t_\mathrm{SNR, \ end}$. Here we roughly assume that the total CR number $N\approx N_0\cdot4\pi R_\mathrm{SNR}^2\cdot 0.1R_\mathrm{SNR}$.  
} \\
}
\end{table*}

In the $20\,\rm{M_\odot}$ scenario described in the previous section (\ref{Subsect:InsideBubble}), the shock front is still located inside the MS progenitor wind bubble. With such a low density of target gas material for $\pi^0$-induced $\gamma$-ray production, the \rm{TeV} emission from the SNR itself cannot be explained in a hadronic emission scenario. As shown in \citet{Ab2011}, to reach the observed $\gamma$-ray emissivity in a hadronic scenario with a still reasonable fraction of SN energy ($\sim0.5\times10^{51}\rm{erg}$) going into cosmic rays, the target gas density has to be of order $1\,\rm{cm^{-3}}$.

It is however possible, also in a similar $\sim 20\,\rm{M_\odot}$ ($\sim 25\,\rm{M_\odot}$) progenitor star scenario like the one above, that the shock has just recently entered the dense shell swept up by the MS wind. This would permit to maintain a high shock speed $v_\mathrm{SNR}>1000\,\rm{km/s}$ and at the same the shock is right now embedded in high circumstellar medium density with $n\gg0.1 \,\rm{cm^{-3}}$, providing thus dense target material for hadronic $\gamma$-rays. Under these well-adjusted conditions, the GeV-TeV spectrum may contain a sizeable fraction of $\pi^0$-induced $\gamma$-rays. To explore such a scenario, we introduced two further scenarios as shown in Table~\ref{table:SNR2}, a $20\,\rm{M}_\odot$ in-shell and a $25\,\rm{M}_\odot$ in-shell scenario. We use the same configuration for these two scenarios as used in the $20\,\rm{M}_\odot$ and $25\,\rm{M}_\odot$ scenarios above except adopting a different initial SN energy and a different MS bubble size with an exponential gas density profile at the shell. This density profile has been used by \citet{Berezhko1713, BerezhkoVelaJr} for the case of RX\,J1713.7-3946, with $n(r) = n_\mathrm{b}+(r/R_\mathrm{b})^{3(\sigma_\mathrm{sh}-1)}n_\mathrm{sh}$, where $n_\mathrm{sh}=\sigma_\mathrm{sh}n_\mathrm{ICM}$ is the maximum density of the shell at the outer boundary of the shell. In our work, we set $\sigma_\mathrm{sh}=5$. The density profile in the RSG bubble and outside of the MS bubble remains the same as in Sect.\,\ref{Subsect:InsideBubble}. 
 
\begin{figure}[h!]
\centering
\begin{tabular}{c } 
\hspace{-0.5cm}
{\includegraphics[width=7.8cm, height=5.cm]{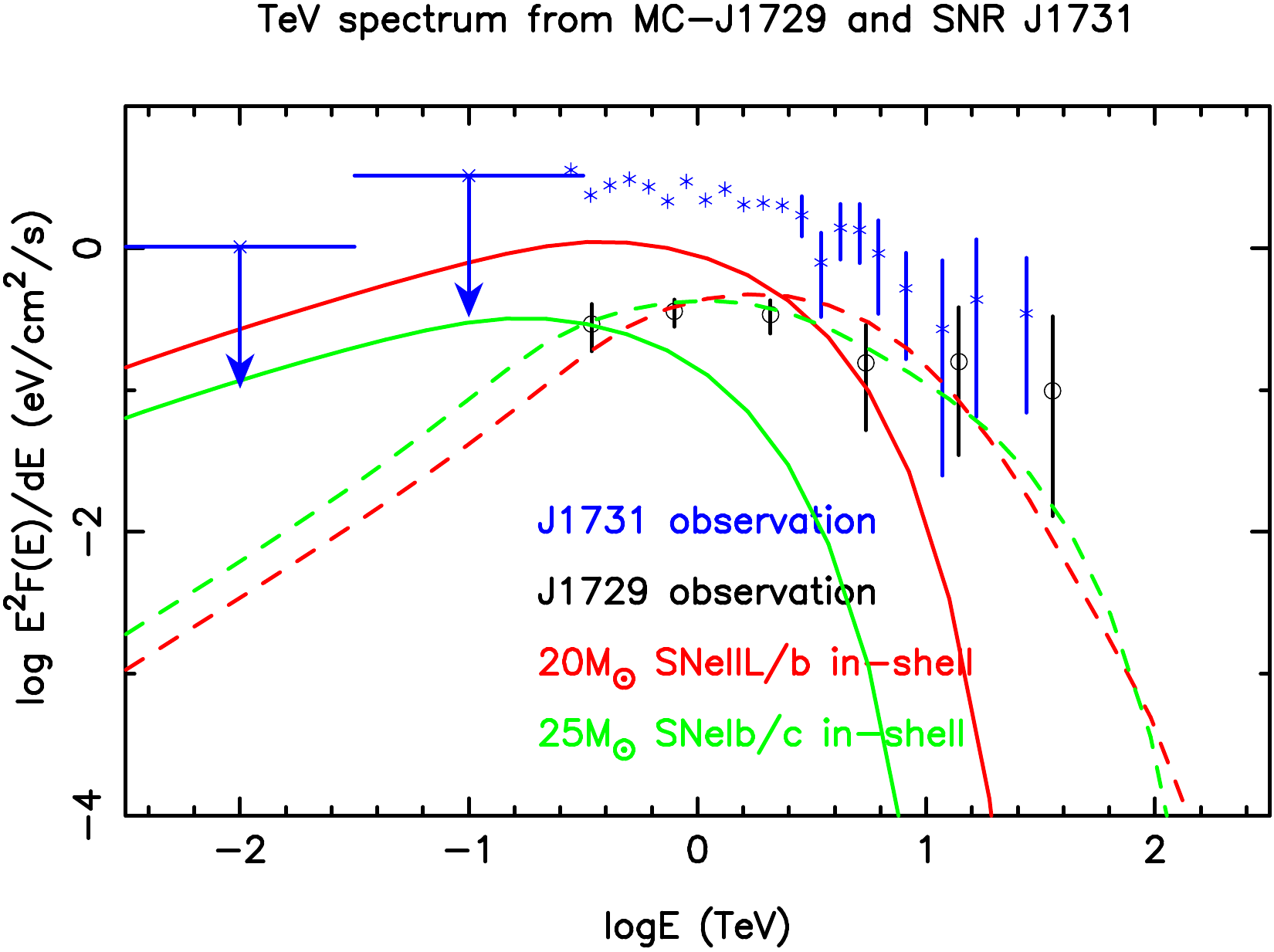} }  
\end{tabular}
\caption{TeV spectra resulting from hadronic CR-gas interactions, from CRs trapped inside the downstream shock region of SNR HESS J1731-347 (solid lines) and from CRs interacting with MC-J1729 (dashed lines). Red/green lines represent the $20\,\rm{M}_\odot$ in-shell/$25\,\rm{M}_\odot$ in-shell scenarios shown in Table.~\ref{table:SNR2}. 
Adopting a galactic standard diffusion coefficient $D_{10}=10^{28}\,\rm{cm^2s^{-1}},\ \delta=0.5$ all over the space, we fitted the spectrum of MC-J1729 (at 30\,pc distance from the center of the SNR). Observational data from \citet{Ab2011} and \citet{Fermi2015} for HESS\,J1731-347 and HESS\,J1729-345 are shown as blue stars and black circles, respectively.
}
\label{fig:spec2}
\end{figure}

As shown in Fig.~\ref{fig:spec2}, we try to fit the spectrum of both HESS J1731-347 and HESS J1729-345 assuming as in Sect.\,\ref{Subsect:InsideBubble} that MC-J1729 is located 30\,pc away from the CCO. The total SN energy for both scenarios are set higher than in Sect.\,\ref{Subsect:InsideBubble} (to $2\times10^{51}$erg) in order to obtain a higher shock speed when the SNR shock hits the MS bubble shell. As shown in Table.~\ref{table:SNR2}, in scenario $20\,\rm{M}_\odot$ in-shell ($25\,\rm{M}_\odot$ in-shell), the MS bubble size is set just such that the SNR keeps a fast enough shock speed of 1150 (1170)\,km/s at present time.

Compared to the $25\,\rm{M}_\odot$ in-shell scenario, more mass has been swept up in the $20\,\rm{M}_\odot$ in-shell scenario at an early phase of the SNR evolution because of the dense RSG wind bubble.
This is equivalent to the SNR having a much higher ejecta mass. Consequently, the shock can enter deeper into the MS bubble shell with high velocity, with a density at the present shock position of $n_{\mathrm{end}} = 0.58\,\rm{cm^{-3}}$, compared to only $n_{\mathrm{end}} = 0.35\,\rm{cm^{-3}}$ for the $25\,\rm{M}_\odot$ in-shell scenario.
Most of the CRs inside the SNR are confined at the acceleration region downstream of the shock. \citet{Zirakashvili2008b} provide the CR density $N_0(E)$ at the shock front as discussed in Sect.\,\ref{Subsect:Evolution}, see also Appendix A. Here, we roughly assume $0.1R_{\mathrm{SNR}}$ \citep{Zirakashvili2012} as the size of the region between shock and contact discontinuity.
The CRs (with density $N_0$) are assumed to be evenly distributed in this region. The total CR energy trapped in this downstream acceleration region can be calculated with $E_\mathrm{CR,sh} = \int dE\cdot E\cdot N_0(E)\cdot4\pi R_\mathrm{SNR}^2\cdot 0.1 R_{\mathrm{SNR}}$. The total swept-up gas at present time can be assumed to be confined inside this region as well, with $\sim60\,\rm{M}_\odot$ / $30\,\rm{M}_\odot$ for scenario $20\,\rm{M}_{\odot}$ in-shell / $25\,\rm{M}_{\odot}$ in-shell. Upstream of the shock, the CRs are mainly confined inside the acceleration region with size $\sim 0.05\,R_{\mathrm{SNR}}$, where the magnetic turbulence is amplified by the CRs. This region has a much lower CR pressure ($\sim$10 times lower) than the downstream region \citep{Zirakashvili2012}. For the $\gamma$-ray spectrum, we therefore did not include the upstream contribution from hadronic interactions, although the total target gas mass here is higher ($\sim $2 times higher in our scenarios) than in the downstream area. In our model, the estimate of the CR density distribution near the shock is consistent with the acceleration theory in \cite{Zirakashvili2008b,Zirakashvili2012}, see e.g. the CR distribution in Fig.~2 (shock speed $3300\,\rm{km/s}$) and Fig.~4 (shock speed $660\,\rm{km/s}$) in \cite{Zirakashvili2012}.

Table~\ref{table:SNR2} lists the parameters for our optimized model scenarios. In order to be compliant with the Fermi-LAT upper limit of the SNR \citep{Fermi2015}, we introduce a very hard CR spectrum at the shock, $N_0\propto E^{-1.5}$. In Fig.\,\ref{fig:spec2}, the resulting $\gamma$-ray spectra are compared to the observational data from HESS J1731-347 and HESS J1729-345. The spectrum of HESS J1729-345 is fitted well in both scenarios with a diffusion coefficient $D_{10}=10^{28}\,\rm{cm^2s^{-1}},\ \delta=0.5$ set all over the space. But the $20\,\rm{M}_\odot$ in-shell/$25\,\rm{M}_\odot$ in-shell scenarios would need about 3/8 times higher CRs density or swept-up mass in order to fit the spectrum of the SNR itself, respectively. For these scenarios, the swept-up gas mass is constrained by the requirement of maintaining a high shock speed, and the downstream CR density is constrained mainly by the total CR energy content generated by the SNR, $E_{\mathrm{CR,end}} + E_{\mathrm{CR,sh}} \lesssim 0.1\,E_{\mathrm{ej}}$. Only when substantially violating these limits and in addition increasing the diffusion coefficient above Galactic standard, an acceptable fit to the data both for HESS J1731-347 and for MC-J1729 could be achieved (not shown).

To summarize the presented shock-``in shell'' scenarios, with the approximate estimates of the CR population and gas densities downstream and upstream of the shock, the corresponding hadronic $\gamma$-ray emission is well below the measured TeV emission from HESS\,J1731-347. This is in agreement with the conclusions by \citet{Fermi2015} concerning the dominant leptonic nature of the $\gamma$-ray emission from ths SNR. There are however additional processes which still may contribute significantly to the $\gamma$-ray emission in direction of HESS\,J1731-347:

\begin{itemize}
\item (Some of the) observed \rm{TeV} emission inside SNR region can be caused by hadronic $\gamma$-ray emission in molecular clouds located far from the SNR but still along the line of sight of the SNR. The corresponding spectral component would resemble the one expected from MC-J1729 (cf. Fig.\ref{fig:spec}), with a peak at high energies. Since the measured TeV morphology of HESS\,J1731-347 is shell-like, a dominant contribution of this effect is not very likely.  
\item Molecular cloud clumps may have survived the MS wind. Being embedded inside the MS bubble, the forward shock of SNR could have passed around them. Such molecular cloud clumps can easily provide $10^{2\sim4}\,\rm{M}_\odot$ target material for hadronic CR-induced $\gamma$-ray emission inside the SNR. Observational evidence may come from the detection of molecular clumps with strong velocity dispersion within the SNR with more detailed CO or CS data, cf.\ e.g.\ to the SNR CTB\,109 \citep{Sasaki2006}.  
\end{itemize}

Recently, \citet{Fukuda2014} suggested that HESS J1731-347 could be associated with gas located in the 3\,kpc arm of our galaxy, using $\rm{^{12}CO}$ data from Nanten and H$\rm{_I}$ data at $-90\,\rm{km/s}$ to $-75\,\rm{km/s}$. In this scenario, the SNR would have a distance to Earth of $\sim5.2\,\rm{kpc}$ and a radius of $\sim25\,\rm{pc}$. A hadronic origin of the TeV $\gamma$-ray emission was argued to be likely under these circumstances. Here, we note that such a setting could also be accommodated for in our model scenarios, if we artificially increase the MS bubble to a size $R_\mathrm{b}>25\,\rm{pc}$. Then, a SNR with $20\,\rm{M_\odot}$ progegenitor mass and a SN kinetic energy of $2\times 10^{51}\rm{erg}$ can maintain a shock speed $\sim$\,$2000\,\rm{km/s}$ after $\sim$\,$8000$ years when the SNR shock expands to $25\,\rm{pc}$.

\subsection{On the isotropic diffusion assumption}
\label{Subsect:Diffusion}

The presented work has shown that the $\gamma$-ray luminosity of HESS J1729-345 can be explained through escaping CRs from the SNR HESS J1731-347 without invoking an order of magnitude lower than Galactic CR diffusion speed in the vicinity of the SNR. MC-J1729 can dominate the TeV emission outside the SNR assuming isotropic diffusion by adopting a reasonable 3-dimensional molecular cloud structure. There is no need to invoke anisotropic CR diffusion \citep{Jokipii1966,Nava2013,Malkov2013,Giacinti2013} to boost the CR density in a particular direction.

Nevertheless, a diffusive transport of CRs is necessary in order to transform an in energy space narrowly-peaked distribution (around $E_{\mathrm{max}}$) of CRs at escape time into a sufficiently widened particle spectrum, that after $\gamma$-ray emission is broadly compatible with the measured TeV spectrum. 

However, the assumption of diffusive CR propagation on scales where cross-field diffusion of particles is not a dominant process may not necessarily be justified. On short scales, high energy CR particles are rather travelling along magnetic flux tubes as bulk movement \citep[e.g.][]{Giacinti2013}, with corresponding decoherence lengths of $\sim 100\,\mathrm{pc}$. 
In the most extreme case, a monoenergetic particle distribution injected near the SNR could be preserved until the CRs hit a nearby molecular cloud. Such a scenario would be incompatible with the TeV $\gamma$-ray spectrum of HESS J1729-345. 
We have to assume that MC-J1729 is irradiated by CRs after an effective diffusion length of 30\,pc, while particles may have traveled possibly on more extended scales before returning to the cloud. 
Therefore, we have also not considered modifications on the $\gamma$-ray emissivity in the line of sight to the observer that result from an anisotropic CR momentum distribution, occuring initially in magnetic flux tube bulk movement as in pure ballistic particle propagation \citep{Prosekin2015}.
The evaluation of possible consequences of this assumption is deferred to future work.

\section{Conclusion}
\label{Sect:DiscussionConclusion}
Through exploring the SNR evolution history with different scenarios ($8\,\rm{M_\odot}$, $15\,\rm{M_\odot}$, $20\,\rm{M_\odot}$, $25\,\rm{M_\odot}$ progenitor masses), we found that the SNR HESS J1731-347 is most likely still expanding inside its progenitor main sequence wind bubble. With $20\,\rm{M_\odot}$ and $25\,\rm{M_\odot}$ progenitor mass scenarios, we successfully reproduce relatively fast shocks $>2000\,\rm{km/s}$ at present time ($\sim6\,\rm{kyr}$ and $\sim3\,\rm{kyr}$ after the SN explosion), which is required by the non-thermal X-ray emission detected from the SNR. 

One of our main goals of this work was to verify whether the TeV emission from the nearby source HESS J1729-345 can be explained with $\gamma$-ray emission from runaway CRs that have been accelerated in HESS J1731-347 and are now illuminating molecular clouds near the SNR. To this extent, a simplified 3-dimensional molecular cloud geometry near the SNR was constructed in accordance with existing $^{12}$CO data. Adopting the CR acceleration model of non-resonant streaming instability from \citet{Zirakashvili2008b}, the CR injection into the surroundings of the SNR was calculated throughout the different simulated SNR evolutions. Diffusion of the CR particles into the inhomogeneous surrounding medium was simulated by means of a Monte Carlo transport code.
By placing HESS J1729-345 30\,pc away from the center of the SNR, its spectrum can be reproduced in scenarios with $20\,\rm{M_\odot}$ and $25\,\rm{M_\odot}$ progenitor masses, with a diffusion coefficient of $D_{10}=10^{28}\,\rm{cm^2s^{-1}}$, $\delta=0.3$, inside the molecular cloud clumps.
 
The following key features of our simulations are basically driven by the young SNR age in our model scenarios:
\begin{itemize}
\item The escaped CRs will concentrate in one energy band at relatively high energies. 
The simulated TeV $\gamma$-ray spectrum of HESS J1729-347 peaks at $\sim1\,\rm{TeV}$. Thus no detection from Fermi is expected in our model. Future HESS-II or CTA observations with high sensitivity around 100\,GeV may also provide key evidence for our model. 
\item The CR density will drop very fast with increasing distance from the SNR.
Inside the $20\,\rm{M_\odot}$ progenitor mass scenario, we also explored the parameter space by adopting different diffusion coefficients at the ICM or inside MC clumps. The observed TeV image can be well explained assuming a galactic diffusion coefficient in the ICM ($D_{10}=10^{28}\,\rm{cm^2s^{-1}},\ \delta=0.3-0.5$). Assuming that HESS J1729-345 is $\sim$\,$30$\,pc away from the center of the SNR, $\gamma$-rays from hadronic interactions from the corresponding molecular cloud can very well dominate the \rm{TeV} emission outside the SNR, when the densest MC region is placed at a larger distance $\sim$\,$100$\,pc from the SNR.
\end{itemize}

Although the GeV upper limits from Fermi-LAT favour a leptonic nature of the $\gamma$-ray emission from the SNR itself, we introduced two additional scenarios ($20\,\rm{M_\odot}$ in-shell and $25\,\rm{M_\odot}$ in-shell) in which the SNR at present time has started to evolve into the progenitor MS bubble shell. Also in these scenarios, the TeV $\gamma$-ray spectrum of HESS J1729-345 can be reproduced with hadronic emission from runaway CRs, when adopting a diffusion coefficient $D_{10}=10^{28}\,\rm{cm^2s^{-1}},\ \delta=0.5$ everywhere. Still, it seems that the hadronic contribution to the SNR $\gamma$-ray emission is not dominating also under such conditions, the expected emission from the downstream region of the shock only provides $\sim30\%$ and $\sim12\%$ of the observed TeV emission, respectively.

\vspace{10ex}

\begin{acknowledgements} 
We thank H. V{\"o}lk for important suggestions on SNR evolution, CR acceleration, and CR diffusion. We thank V. Zirakashvili for extensive support on CR acceleration theory. We thank S. Ostapchenko providing the proton-proton collision cross section table. The work was supported by the German Science Foundation (DFG) in the DFG priority program SPP 1573 through grant PU 308/1-1.
\end{acknowledgements}

\begin{appendix} 
\section{The acceleration theory of \citet{Zirakashvili2008b}}

The momentum distribution of CRs at the shock front can be written as:
\begin{equation}
N_0(p)=\frac{\eta_\mathrm{esc} \rho u_1^2}{8\pi cI} p^{-\gamma_\mathrm{s}}p^{\gamma_\mathrm{s}-4}_\mathrm{m} n_0(\frac{p}{p_\mathrm{m}})  ;
\end{equation}
$u_1=v_\mathrm{SNR}$ is the shock speed, 
$\rho$ is the density of the ambient gas, $p$ ($p_\mathrm{max}$) is the momentum (maximum momentum) of the CRs.
$I$ is a normalisation factor, for $\gamma_\mathrm{s}=4$ (as used in Sect.\,\ref{Subsect:InsideBubble}) $I = 1/4$.
$\eta_\mathrm{esc} = \frac{F_\mathrm{esc}}{1/2\rho u_1^3} $ 
is the ratio between the energy flux of escaping CRs and the kinetic energy flux of the upstream medium that is approaching the shock,
and 
\begin{equation}
n_0 (s)= exp[ -\gamma_\mathrm{s} \int_0^s \frac{ds_1/s_1}{exp(s_1^{-2})-1} ] . 
\label{equ:n_0}
\end{equation}

Beyond an absorption boundary, CR particles are assumed to travel freely into space. The absorption boundary is set to $L = t_\mathrm{SNR} v_\mathrm{SNR}$, where $L$ is measured from the shock position $R_\mathrm{SNR}$ in radial direction outwards.
The flux of escaping particles at the absorption boundary can be written as follows:
\begin{equation}
J(p)= u_1N_0(p)/(e^{p_\mathrm{m}^2/p^2}-1).
\end{equation}

In \citet{Zirakashvili2008b} and the presented work, the momentum distribution of particles is used, which can be transformed to an energy distribution through
$f(E)dE=4\pi p^2f(p) dp$, $E=pc$.

To obtain the escape energy $p_\mathrm{m}c$, 
one needs to know the size of the absorption boundary,
the shock speed $u_1$, the density of the incoming gas from upstream $n_\mathrm{H}$, and the amplified magnetic field $B_\mathrm{r}$. 
\citet{Zirakashvili2008b} provide
an approximated analytical solution:  
\begin{equation}
p_\mathrm{m}c=\frac {\eta _\mathrm{esc}qu_1^2B_0L}{4cV_\mathrm{A}}
\left\{ \begin{array}{ll}
\ln ^{-1}\left( \frac {2B_0u_1^2}{B_\mathrm{b}u_*^2}\right) , u_1<u_*\\
\left[ \ln \left( \frac {2B_0}{B_\mathrm{b}}\right) -1 +\left( 2\frac
{u_1^4}{u_*^4}-1\right) ^{1/4} \right] ^{-1}, 
u_1>u_*
\end{array} \right.
\end{equation}
where the velocity $u_*=(24\pi cV_\mathrm{A}^3/\eta _\mathrm{esc}^2)^{1/4}$, and  the Alfv\'en velocity $V_\mathrm{A}=B_0/\sqrt{4\pi \rho }= 2.18\,\mathrm{km\,s^{-1}} (\frac{n_\mathrm{H}}{1\,H\,\mathrm{cm^{-3}}})^{-1/2}(\frac{B}{1\,\mathrm{\mu G}})$.  $B_0$ is the original unamplified
magnetic field in the upstream region, and
$B_\mathrm{b}$ is the initial value of magnetic fluctuations in the upstream region, here set to $ln(2B_0/B_\mathrm{b})=5$.

One important caveat is that in the CR acceleration model as elaborated in \citet{Zirakashvili2008b}, nonresonant instabilities only become efficient when shock speed is high enough, $v_\mathrm{sh} > (1340\,\mathrm{km/s}) (V_\mathrm{A}/10\,\mathrm{km/s})^{2/3} (\eta_\mathrm{esc}/0.05)^{-1/3} $. 

\section{The SNR evolution histories for all modelled scenarios}

\begin{figure}[h!]
\begin{tabular}{c}
{\includegraphics[width=7.8cm, height=5.2cm]{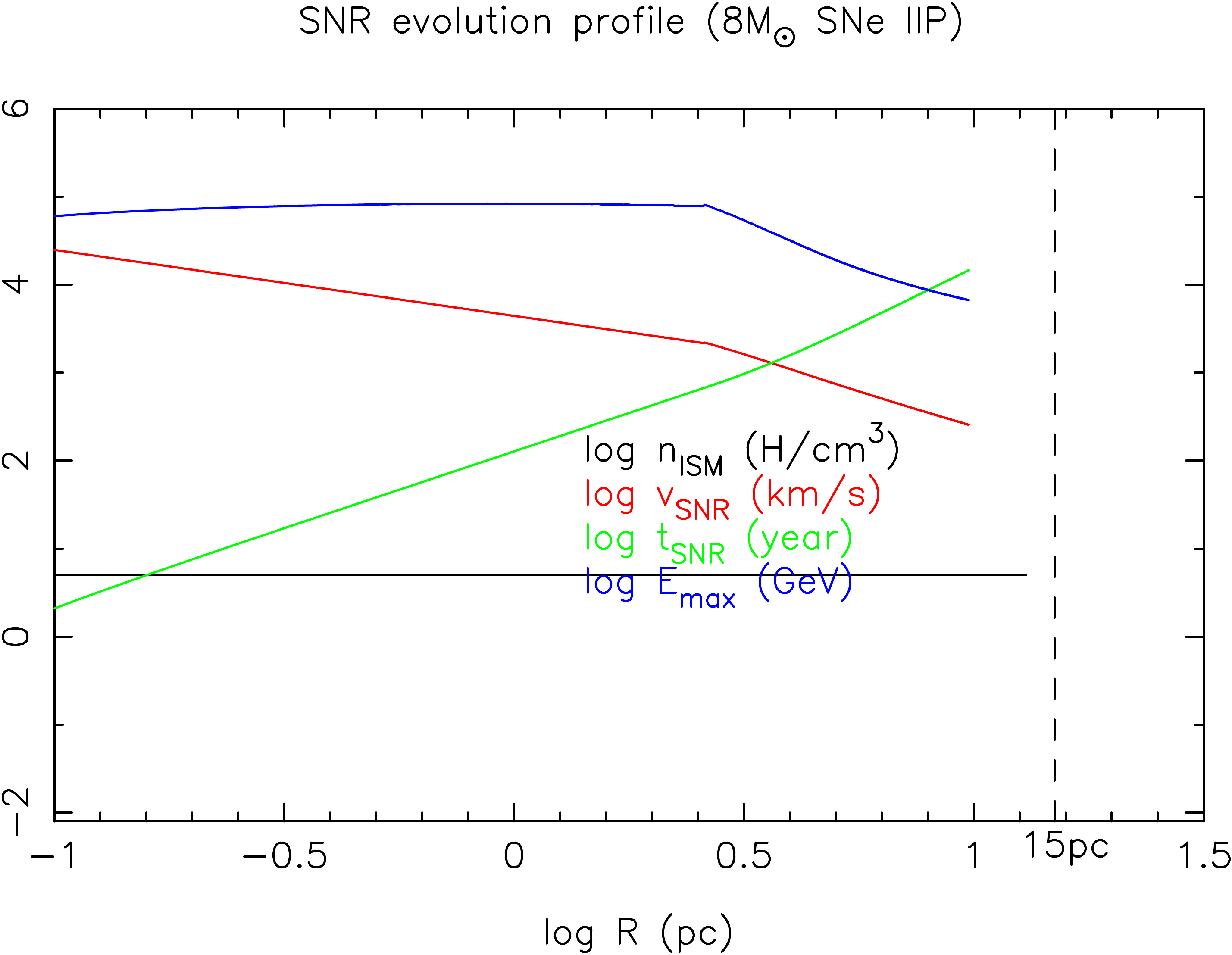} } \\
{\includegraphics[width=7.8cm, height=5.2cm]{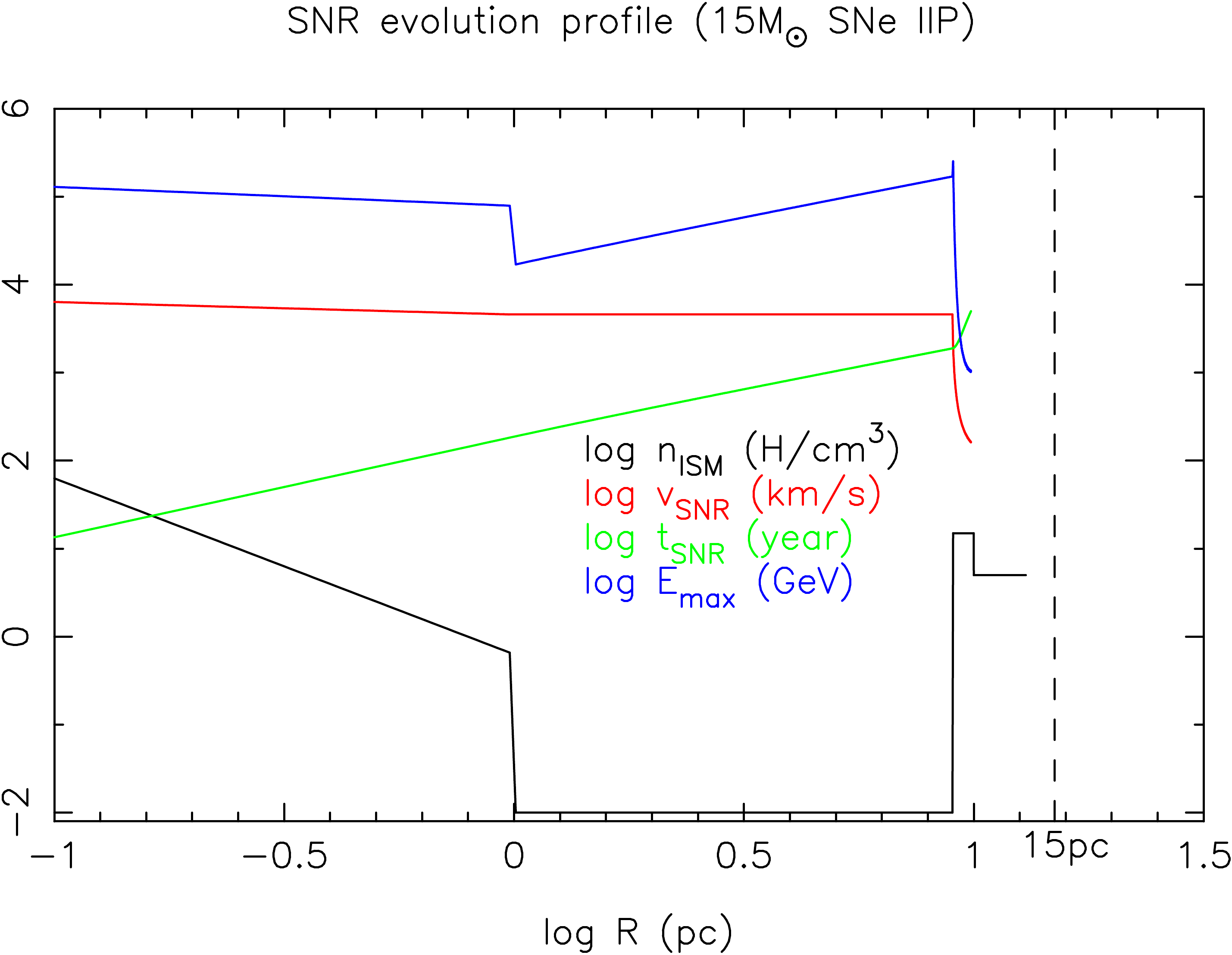} } \\
{\includegraphics[width=7.8cm, height=5.2cm]{SNR_evolution_20M} } \\
{\includegraphics[width=7.8cm, height=5.2cm]{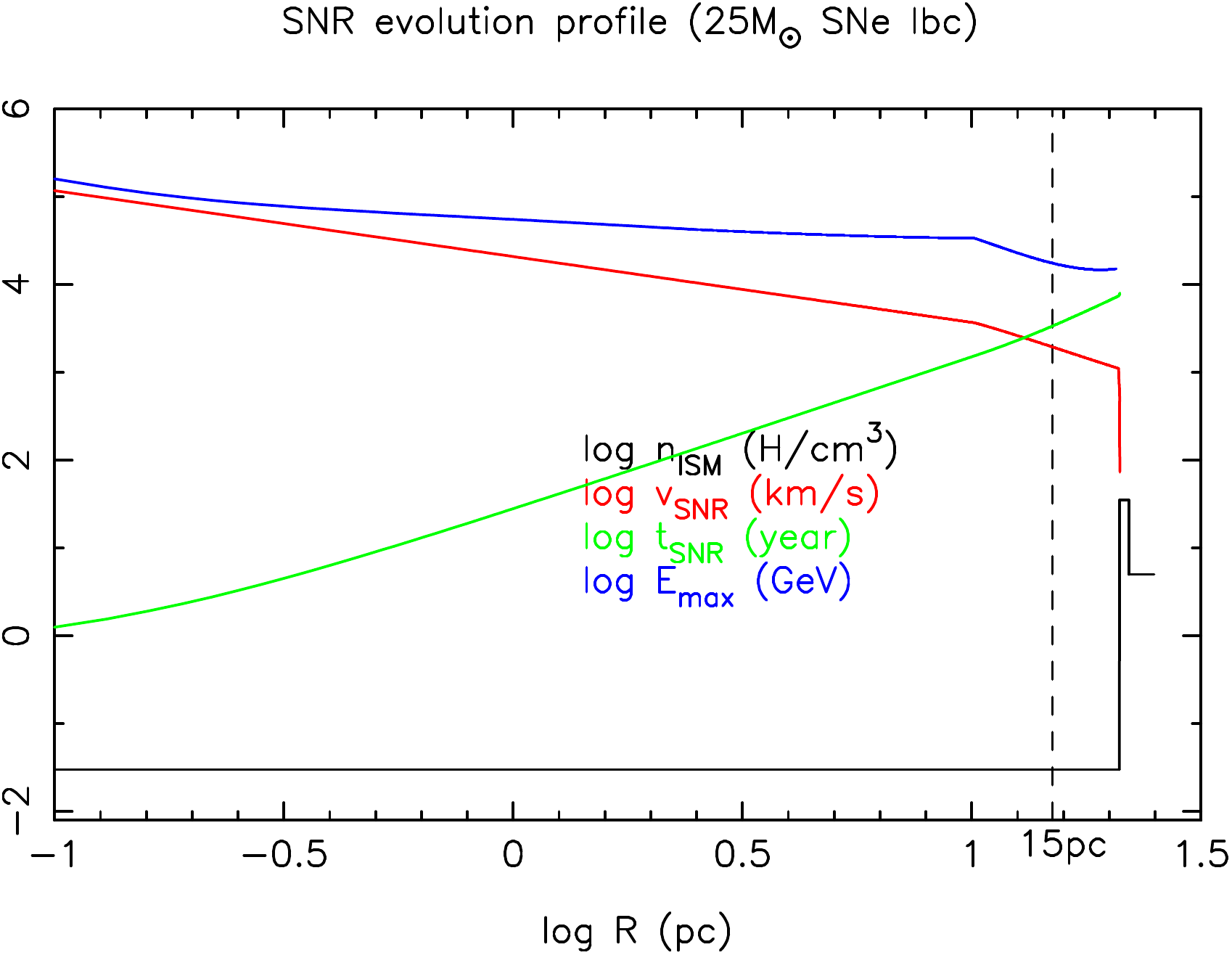} } \\
\end{tabular}
\caption{The SNR evolution profile with the shock velocity $v_{\mathrm{SNR}}$ (red), the age of the SNR $t_{\mathrm{SNR}}$ (green), and the CR particle escape energy $E_{\mathrm{max}}$ (blue) plotted against the SNR radius. Also shown is the density of the circumstellar medium $n_{\mathrm{ISM}}$ (black) out to 25\,pc distance.
From top to bottom are four different scenarios: $8\,\rm{M_\odot}$ SN IIP, $15\,\rm{M_\odot}$ SN IIP, $20\,\rm{M_\odot}$ SN IIL/b, and $25\,\rm{M_\odot} $ SN Ib/c.  
The current SNR size (15\,pc) is marked with a dashed line. The time, velocity and escape energy profiles are cut off when the shock speed is too low ($v_\mathrm{SNR}\ll1000\,\rm{km/s}$) for applying the prescriptions of \cite{Zirakashvili2008b} to compute a CR particle escape energy. See detailed parameters in Table 1.  }
\label{fig:SNRs}
\end{figure}

\begin{figure}[h!]
\begin{tabular}{c}
\hspace{-.cm}
{\includegraphics[width=7.8cm, height=5.cm]{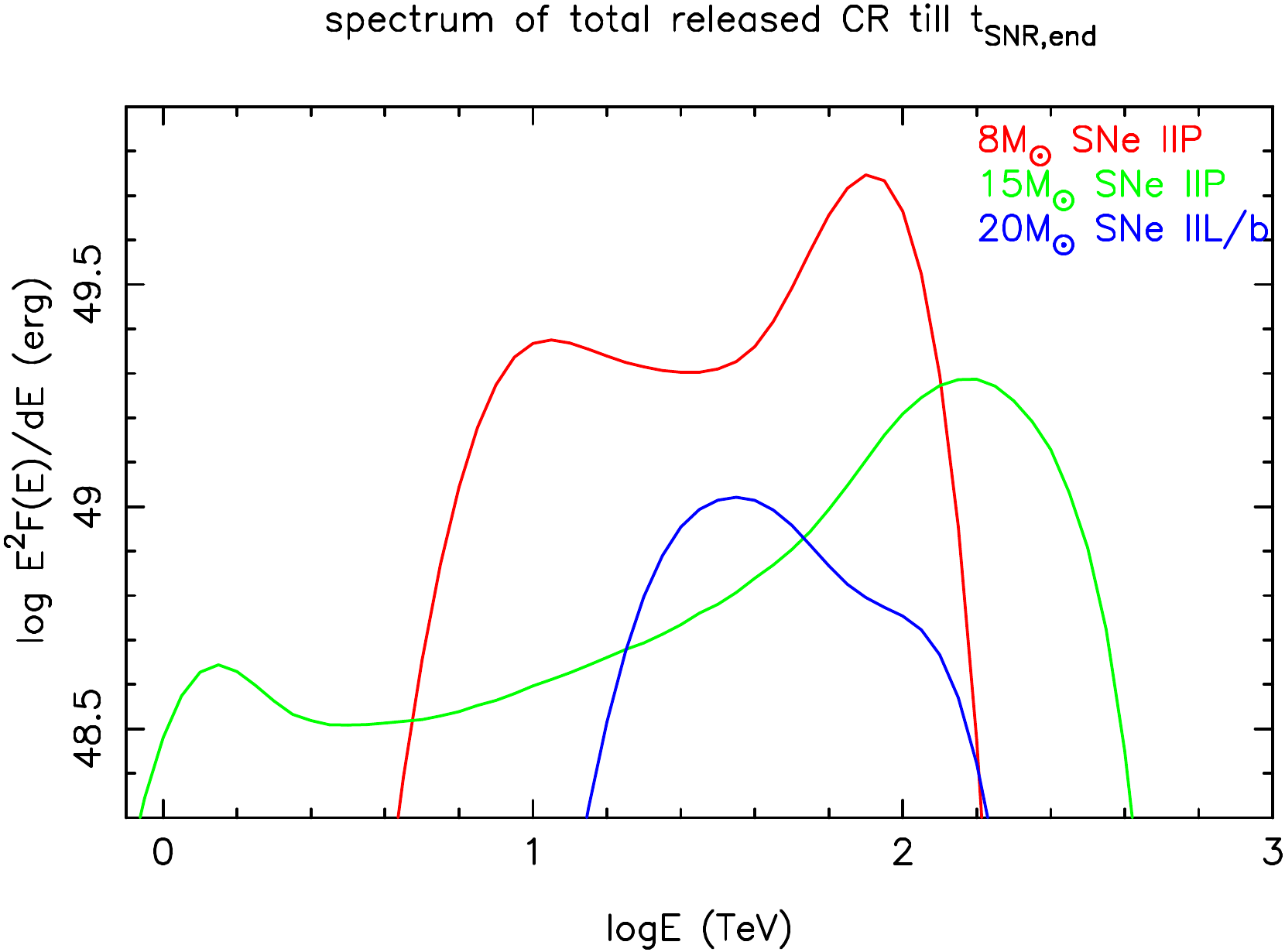} } \\
{\includegraphics[width=7.8cm, height=5.cm]{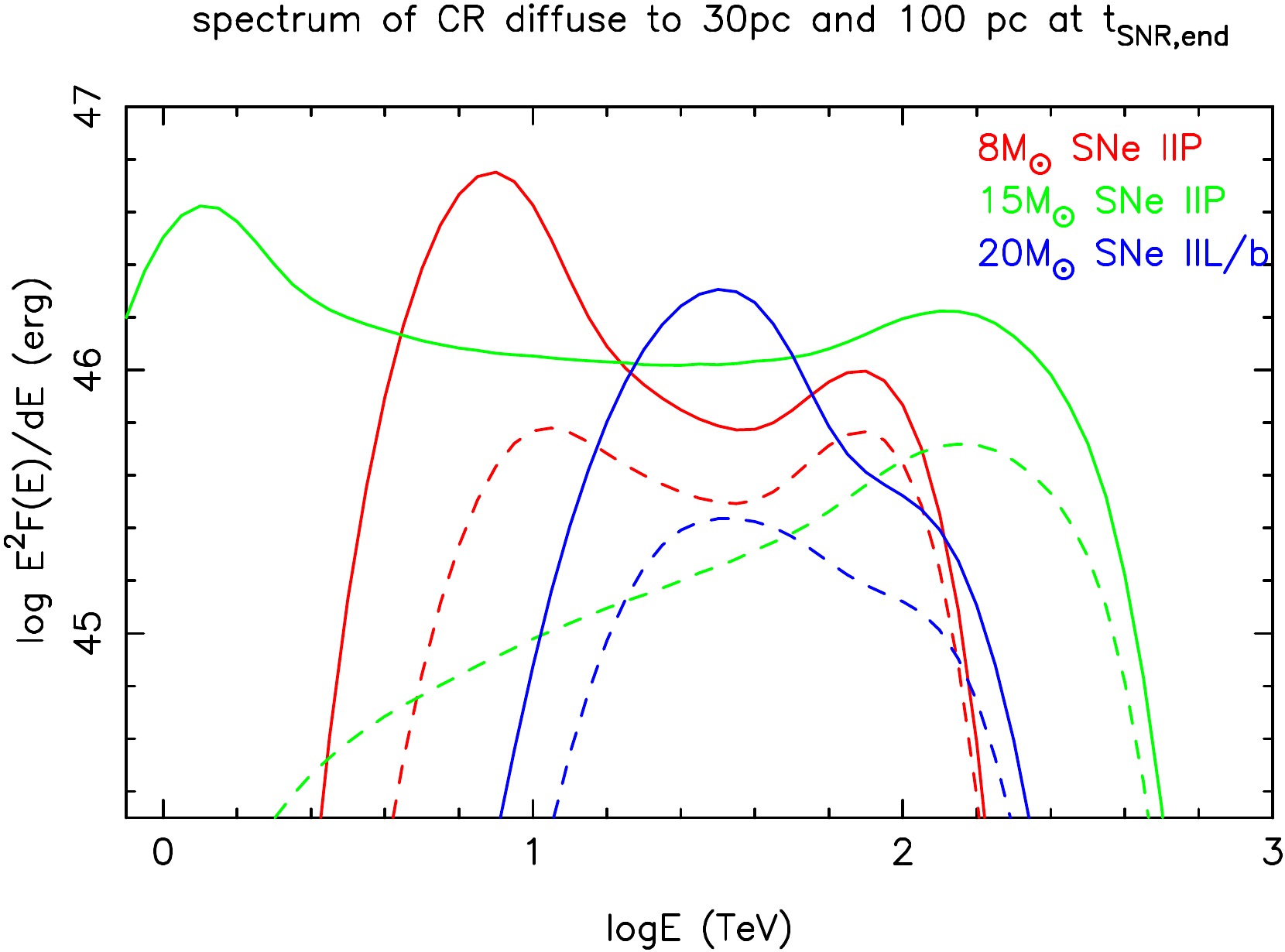} } \\
\end{tabular}
\caption{{\it Top panel:} The particle spectra of all CRs released from the SNR, integrated from the time of the SN explosion to $t_\mathrm{SNR, end}$. The SNR ages at $t_\mathrm{SNR, end}$ are $15.5\,\rm{kyr}$, $5.8\,\rm{kyr}$, $6.1\,\rm{kyr}$, and $2.9\,\rm{kyr}$, respectively, corresponding to the $8\,\rm{M_\odot}$ SN IIP (red),  $15\,\rm{M_\odot}$ SN IIP (green), $20\,\rm{M_\odot}$ SN IIL/b (blue), and $25\,\rm{M_\odot}$ SN Ib/c (cyan) scenarios.  {\it Bottom panel:} The particle spectra of CRs that have diffused in a homogeneous environment from the SNR to 30\,pc (solid lines) and 100\,pc (dashed lines) distance. These spectra are obtained by multiplying the CR density at 30\,pc and 100\,pc with the volume of MC-J1729 ($4/3\pi R^3,\ R=7.8\,\rm{pc}$) and the volume of MC-core ($R=7.2\,\rm{pc}$), respectively. }
\label{fig:CR}
\end{figure}

\begin{figure}[h!]
\begin{tabular}{c}
\hspace{-.cm}
{\includegraphics[width=7.8cm, height=5.cm]{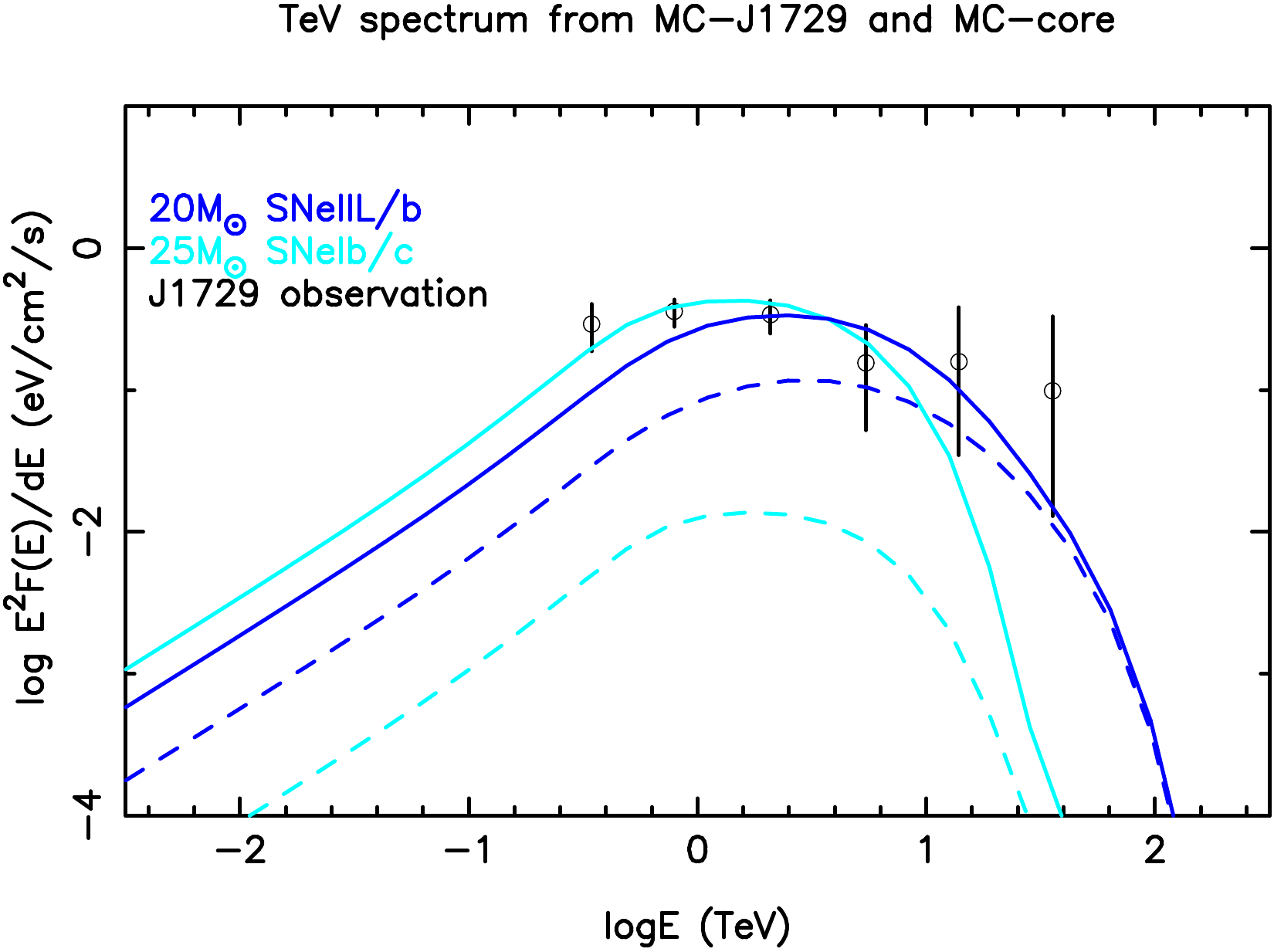} } 
\end{tabular}
\caption{Simulated $\gamma$-ray spectra from hadronic collisions of CRs inside MC-J1729 (solid lines) and MC-core (dashed lines).
The spectra are obtained by multiplication of the simulated CR density at 30\,pc (100\,pc) with the total mass of MC-J1729 (MC-core), for the $20\,\rm{M_\odot}$ SN IIL/b (blue) and the $25\,\rm{M_\odot}$ SN Ib/c (cyan) scenarios.
Observational data from \citet{Ab2011} for HESS J1729-345 are shown as black circles. }
\label{fig:spec_analytical}
\end{figure}

The SNR evolution histories of all four SN scenarios discussed
in Sect.\,\ref{Subsect:InsideBubble} are represented in  Fig.~\ref{fig:SNRs}, the parameters used here are the ones from Table 1. Our SNR models need to evolve inside different regions with significant density changes of the circumstellar medium. For the Sedov phase, the ``thin-shell'' approximation solution can naturally cover those density changes, but for the ejecta-dominated phase, each scenario is treated individually.

\begin{itemize}
\item In the $8\,\rm{M_\odot}$ scenario, the SNR radius first follows $R_\mathrm{SNR} \propto t^{4/7}$ during the ejecta-dominated phase; after the SNR has swept up $\sim1.5\,M_\mathrm{ej}$ of ICM at around 800 years, it enters the Sedov phase. 
\item In the $15\,\rm{M_\odot}$ scenario, inside the RSG bubble the SNR evolution follows $R_\mathrm{SNR} \propto t^{7/8}$; once the SNR reaches the empty MS wind bubble, we simply assume that the shock velocity remains constant. Right after the SNR reaches the MS wind bubble shell at $\sim 2000$\,years, the SNR quickly enters the Sedov phase in less than 100 years.
\item In the $20\,\rm{M_\odot}$ scenario, the SNR radius first follows $R_\mathrm{SNR} \propto t^{7/8} $ during the ejecta-dominates phase. At $\sim 30$\,years, when the SNR is still inside the RSG bubble, the SNR has swept up $\sim0.4\,M_\mathrm{ej}$ and enters the Sedov phase.
\item In the $25\,\rm{M_\odot}$ scenario, the SNR radius first follows $R_\mathrm{SNR} \propto t^{4/7} $ during the ejecta-dominated phase. At $\sim 1700$\,years, the SNR has swept up $\sim1.6\,M_\mathrm{ej}$ and enters the Sedov phase.
\end{itemize}

In Fig.~\ref{fig:CR} and Fig.~\ref{fig:spec_analytical}, a homogeneous diffusion coefficient ($D_{10}=10^{28}\,\rm{cm^2/s}, \ \delta=0.3$) is adopted all over the place, which means that we can directly obtain the density of the diffusing CRs at 30\,pc and 100\,pc distance from the SNR at present time $t_\mathrm{end}$ through an analytical diffusion solution. 
To roughly estimate the final simulated TeV spectrum as shown in Fig.~\ref{fig:spec_analytical}, the CR density at the clouds is multiplied with the masses of MC-J1729 and MC-core, respectively.

\end{appendix}
\end{document}